\documentclass[journal]{IEEEtran}
\ifCLASSINFOpdf
\else
   \usepackage{graphicx}
   \graphicspath{{../eps/}}
   \DeclareGraphicsExtensions{.eps}
\fi

\usepackage{amsmath}
\usepackage{amsfonts,amssymb}
\usepackage{amssymb}
\usepackage{wrapfig}
\usepackage{psfrag}
\usepackage{epstopdf}
\usepackage{cite}
\usepackage{graphicx}
\usepackage{subfigure}
\usepackage{threeparttable}
\usepackage{cases}
\usepackage{subeqnarray}
\usepackage{color}
\usepackage{underscore}
\usepackage{verbatim}
\usepackage{bm}
\usepackage{stfloats}
\usepackage{xpatch}

\newtheorem{algorithm}{Algorithm}
\newtheorem{proposition}{Proposition}

\usepackage{algorithm}
\usepackage{algorithmic}

\makeatletter
\def\changeBibColor#1{%
  \in@{#1}{}
  \ifin@\color{red}\else\normalcolor\fi
}
\xpatchcmd\@bibitem
  {\item}
  {\changeBibColor{#1}\item}
  {}{\fail}
\xpatchcmd\@lbibitem
  {\item}
  {\changeBibColor{#2}\item}
  {}{\fail}
\makeatother

\begin{document}
\title{Delay-Doppler Alignment Modulation for Spatially Sparse Massive MIMO Communication}
%
%
%

\author{Haiquan~Lu
        and
        Yong~Zeng,~\IEEEmembership{Senior Member,~IEEE}
\thanks{This work was supported in part by the National Key R\&D Program of China with Grant number 2019YFB1803400, and in part by the Fundamental Research Funds for the Central Universities 2242022k60004. (\emph{Corresponding author: Yong Zeng.}) }
\thanks{Haiquan Lu and Yong Zeng are with the National Mobile Communications Research Laboratory and Frontiers Science Center for Mobile Information Communication and Security, Southeast University, Nanjing 210096, China, and also with the Purple Mountain Laboratories, Nanjing 211111, China (e-mail: \{haiquanlu, yong_zeng\}@seu.edu.cn). }
}

\maketitle

\begin{abstract}
\emph{Delay alignment modulation} (DAM) is an emerging technique for achieving inter-symbol interference (ISI)-free wideband communications using spatial-delay processing, without relying on channel equalization or multi-carrier transmission. However, existing works on DAM only consider multiple-input single-output (MISO) communication systems and assume time-invariant channels. In this paper, by extending DAM to time-variant frequency-selective multiple-input multiple-output (MIMO) channels, we propose a novel technique termed \emph{delay-Doppler alignment modulation} (DDAM). Specifically, by leveraging \emph{delay-Doppler compensation} and \emph{path-based beamforming}, the Doppler effect of each multi-path can be eliminated and all multi-path signal components may reach the receiver concurrently and constructively. We first show that by applying path-based zero-forcing (ZF) precoding and receive combining, DDAM can transform the original time-variant frequency-selective channels into time-invariant ISI-free channels. The necessary and/or sufficient conditions to achieve such a transformation are derived. Then an asymptotic analysis is provided by showing that when the number of base station (BS) antennas is much larger than that of channel paths, DDAM enables time-invariant ISI-free channels with the simple delay-Doppler compensation and path-based maximal-ratio transmission (MRT) beamforming. Furthermore, for the general DDAM design with some tolerable ISI, the path-based transmit precoding and receive combining matrices are optimized to maximize the spectral efficiency. Numerical results are provided to compare the proposed DDAM technique with various benchmarking schemes, including MIMO-orthogonal time frequency space (OTFS), MIMO-orthogonal frequency-division multiplexing (OFDM) without or with carrier frequency offset (CFO) compensation, and beam alignment along the dominant path.
\end{abstract}


\IEEEpeerreviewmaketitle
\section{Introduction}\label{sectionIntroduction}
 By exploiting the high spatial resolution of large antenna arrays \cite{bjornson2019massive,lu2022communicating,sanguinetti2020toward} and the multi-path sparsity of millimeter wave (mmWave) and Terahertz (THz) channels \cite{akdeniz2014millimeter,zeng2016millimeter,rappaport2019wireless,chen2021hybrid}, \emph{delay alignment modulation} (DAM) was recently proposed in \cite{lu2022delay,lu2023Manipulating} to resolve the inter-symbol interference (ISI) issue with spatial-delay processing, without resorting to the conventional channel equalization or multi-carrier transmission. Specifically, by judiciously performing \emph{delay pre-/post-compensation} and \emph{path-based beamforming}, DAM renders it possible to manipulate the channel delay spread for time-dispersive channels, so that all multi-path signal components may reach the receiver concurrently and constructively, rather than causing the detrimental ISI. Moreover, DAM establishes an unified framework to achieve ISI-free communication with single- or multi-carrier transmissions via the simple spatial-delay processing \cite{lu2023Manipulating}. The research on DAM is still in its infancy, and some preliminary works are reported in  \cite{lu2022delay,lu2023Manipulating,ding2022channel,lu2023delayIRS,wang2022multi,xiao2022integrated,wang2022bidirectional}.
 In \cite{lu2023Manipulating}, a novel DAM-orthogonal frequency-division multiplexing (OFDM) technique involving joint time- and frequency-domain beamforming was proposed, showing superiority to OFDM in terms of spectral efficiency, bit error rate (BER), and peak-to-average-power ratio (PAPR). By leveraging the block sparsity of mmWave channels, an efficient channel estimation method for DAM was studied in \cite{ding2022channel}. The extensions of DAM to multiple-intelligent reflecting surfaces (IRSs) and multi-user communications were respectively investigated in \cite{lu2023delayIRS} and \cite{wang2022multi}. Besides, DAM-based integrated sensing and communication (ISAC) can be found in \cite{xiao2022integrated} and \cite{wang2022bidirectional}.

 Note that DAM differs from other existing ISI-mitigation techniques like time-domain equalization (TEQ) or OFDM in the following aspects. Firstly, by exploiting the abundant spatial dimension and multi-path sparsity, DAM achieves ISI mitigation via the simple spatial-delay processing, while the TEQ techniques such as time-reversal (TR) \cite{emami2004matched,han2012time,Why2016Chen} and channel shortening \cite{melsa1996impulse,martin2005unification} mainly concentrate on the time-domain processing. Specifically, by regarding the channel as a natural matched filter, TR either suppresses the ISI via the rate back-off technique \cite{emami2004matched,han2012time} or asymptotically eliminates it if infinite number of base station (BS) antennas is available \cite{pitarokoilis2012optimality}. By contrast, DAM is capable of completely eliminating the detrimental ISI by means of flexible delay compensation and path-based beamforming, as long as the number of BS antennas is no smaller than that of the multi-paths. Secondly, different from the dominant multi-carrier OFDM technology that requires a cyclic prefix (CP) for each OFDM symbol to avoid inter-block interference, only one guard interval is needed for DAM within each channel coherence block \cite{lu2022delay,lu2023Manipulating}, and thus higher spectral efficiency is anticipated. Moreover, DAM is free from the practical issues for OFDM, e.g., high PAPR, severe out-of-band (OOB) emission, and sensitivity to carrier frequency offset (CFO). It is worth mentioning that though the relevant tapped delay line technique can be found in \cite{taniguchi2004maximum}, it did not exploit the super spatial resolution and spatially sparse property of mmWave massive multiple-input multiple-output (MIMO) systems. Besides, the idea of path-based delay-compensation was proposed in our previous works \cite{zeng2016millimeter,zeng2018multi}, but they were restricted to communication systems with the special lens antenna arrays.

 Besides frequency selectivity, practical wireless channels are also time-variant due to the relative motion of the transmitter, receiver, and scatterers. In particular, the large Doppler spread caused by the high mobility may destroy the orthogonality among OFDM sub-carriers, which results in inter-carrier interference (ICI). Extensive efforts have been devoted to tackle this issue. For example, an ICI self-cancellation scheme was proposed in \cite{zhao2001intercarrier}, where one data symbol is modulated onto a group of adjacent sub-carriers with predefined weighting coefficients, so that the ICI can be approximately ``self-cancelled''. In \cite{tan2004reduced}, the pulse shaping technique, also known as windowing, was adopted to restrict the ICI power. Besides, various linear or nonlinear channel equalization techniques were studied to suppress the ICI by compensating for the channel distortion \cite{cai2003bounding,schniter2004low,molisch2007iterative}. By exploiting the high spatial resolution of massive MIMO, an angle-domain Doppler compensation technique was proposed in \cite{guo2019high}, for which the dominant Doppler frequency over each matched filter beamforming-based branch is pre-compensated to reduce the Doppler spread. The authors in \cite{hadani2017orthogonal} proposed a novel orthogonal time frequency space (OTFS) modulation scheme, where the doubly selective channel is converted into quasi-time-invariant channel in the delay-Doppler domain, rendering it attractive for high-mobility scenarios. However, the aforementioned techniques either incur performance loss or entail complicated signal processing. For instance, the ICI self-cancellation scheme improves the carrier-to-interference power ratio at the cost of spectral efficiency, and pulse shaping achieves better ICI reduction by using larger roll-off factor, i.e., occupying larger excess bandwidth. The signal processing complexity of equalization becomes prohibitive for the channel with the severe delay/Doppler spread. Note that for multi-antenna OFDM and OTFS systems, channel state information (CSI) is also needed to implement the multi-antenna beamforming. In particular, thanks to the spatial sparsity of mmWave/THz communications, the time-domain channel estimation via estimating the propagation delay and Doppler shift per path, helps reduce the training overhead since fewer channel parameters need to be estimated \cite{liu2014channel}.

 On the other hand, existing works on DAM \cite{lu2022delay,lu2023Manipulating,lu2023delayIRS,ding2022channel,wang2022multi,xiao2022integrated}
 only consider multiple-input single-output (MISO) systems and assume  time-invariant frequency-selective channels. In this paper, by considering the more challenging time-frequency doubly selective MIMO channels, we extend DAM by proposing a Doppler-ISI double mitigation technique termed \emph{delay-Doppler alignment modulation} (DDAM) for high-frequency systems, e.g., mmWave or THz communications. The key idea of DDAM lies in both \emph{delay-Doppler pre-/post-compensation} and \emph{path-based beamforming}, which exploits both the high spatial resolution of large antenna arrays and multi-path sparsity of mmWave/THz channels. To be specific, by judiciously performing delay-Doppler pre-/post-compensation matching the respective multi-paths, and in conjunction with path-based beamforming at the BS and the user equipment (UE), the Doppler effect of each multi-path can be eliminated and all multi-path signal components may reach the UE concurrently and constructively. As such, the time-variant frequency-selective channels can be transformed into time-invariant ISI-free channels, without requiring the sophisticated channel equalization or multi-carrier transmission. DDAM technique has been investigated for the uplink single-input multiple-output (SIMO) ISAC system in our previous work \cite{xiao2022exploiting}. It is worth mentioning that DDAM shares the similarity as \cite{song2019efficient} and \cite{miretti2022little} in advocating the single-carrier transmission for mmWave and sub-THz communications. However, they also have quite important differences. In particular, different from the works \cite{song2019efficient} and \cite{miretti2022little} that only use the single dominant channel path, DDAM tries to make full use of all the significant multi-paths. Besides, while techniques such as CFO/timing offset (TO) pre-compensations are used in OFDM systems, the main novelty of DDAM lies in the exploitation of high spatial dimension of large antenna arrays and multi-path sparsity. Moreover, though DDAM is initially motivated by channel sparsity, it is also applicable to non-sparse channels, since as long as the antenna array is large enough, the multi-paths can be well separated. Besides, even when the multi-paths cannot be fully separated, we may apply the generic DAM technique that tolerates some residual ISI \cite{lu2023Manipulating}.
 Our main contributions are summarized as follows:
 \begin{itemize}[\IEEEsetlabelwidth{12)}]
 \item Firstly, we present the key idea and the transceiver architecture of MIMO DDAM, followed by the derivation of sufficient and/or necessary conditions for transforming the time-frequency doubly selective channels into time-invariant ISI-free channels, which achieves the Doppler-ISI dual mitigation, without resorting to the conventional channel equalization or multi-carrier transmissions. Specifically, for a MIMO system with $M_t$ antennas at the BS and $M_r$ antennas at the UE, which sends $N_s$ data streams over a channel with $L$ multi-paths, a sufficient condition to achieve the above transformation is ${M_t} \ge \left( {L - 1} \right){M_r} + {N_s}$. Furthermore, based on the obtained equivalent channel of DDAM transmission, the optimal path-based ZF precoding and receive combining matrices are derived in closed-form.
 \item Secondly, for the asymptotic case when $M_t \gg L$, it is shown that DDAM is able to achieve the time-invariant ISI-free communications while still preserving all the multi-path signal components, with the simple delay-Doppler compensation and path-based maximal-ratio transmission (MRT) beamforming. Furthermore, when some residual ISI is tolerable, the spectral efficiency of DDAM is maximized by jointly optimizing the path-based transmit precoding and receive combining matrices. By exploiting the double timescales for the composite time-frequency channel and the state of the individual multi-paths (e.g., angle of departures/arrivals (AoDs/AoAs), delays and Doppler frequencies), the mean-square error (MSE)-based method is applied for solving the spectral efficiency maximization problem within each \emph{path invariant time}, for which the state of individual multi-paths is nearly constant.
 \item Lastly, we consider the benchmarking schemes of MIMO-OFDM considering ICI and MIMO-OTFS. Furthermore, the impacts of guard interval on the spectral efficiency are studied for OFDM, OTFS, and DDAM. One important finding is that to avoid the ISI across different path invariant blocks, DDAM only requires one guard interval for each path invariant block rather than for each channel coherence block \cite{lu2022delay,lu2023delayIRS}, which further reduces the guard interval overhead of DDAM.
 \end{itemize}

 The rest of this paper is organized as follows. Section \ref{sectionSystemModel} presents the system model and the key idea of DDAM, together with the derivation of sufficient and/or necessary conditions for transforming time-frequency doubly selective channels into time-invariant ISI-free channels. Section \ref{sectionAsymptoticAnalysis} provides the asymptotic analysis for DDAM. In Section \ref{sectionPathBasedZFPrecoding}, we propose the path-based ZF precoding and receive combining design towards time-invariant ISI-free DDAM communication. In Section \ref{sectionJointPrecodingCombining}, the joint path-based precoding and combining optimization tailored for DDAM is studied by tolerating some residual ISI. Section \ref{sectionMIMOOFDMCommunication} considers the benchmarking schemes of MIMO-OFDM and MIMO-OTFS. Numerical results are presented in Section \ref{sectionNumericalResults}. Finally, Section \ref{sectionConclusion} concludes this paper.

 \emph{Notations:} Scalars are denoted by italic letters. Vectors and matrices are denoted by bold-face lower- and upper-case letters, respectively. ${{\mathbb{C}}^{M \times N}}$ and ${{\mathbb{R}}^{M \times N}}$ represent the space of $M \times N$ complex-valued and real-valued matrices, respectively. For an arbitrary-size matrix ${\bf A}$, its complex conjugate, transpose, and Hermitian transpose are denoted by ${\bf A}^*$, ${\bf A}^T$, ${\bf A}^H$, respectively; ${\rm{rank}}\left( {\bf{A}} \right)$ and ${\left\| {\bf{A}} \right\|_F}$ denote the rank and Frobenius norm of ${\bf A}$, respectively; and ${\rm{vec}}\left( {\bf{A}} \right)$ denotes the column-wise vectorization of ${\bf{A}}$. For a vector $\bf{x}$, $\left\| {\bf{x}} \right\|$ denotes its Euclidean norm. For a real number $x$, $\left\lceil x \right\rceil $ denotes the smallest integer that is greater than or equal to $x$. The notations $\circledast$ and $\otimes$ represent the linear convolution and the Kronecker product operations, respectively. The distribution of a circularly symmetric complex Gaussian (CSCG) random vector with mean $\bf{x}$ and covariance matrix $\bf{\Sigma}$ is denoted by ${\cal CN}\left( {\bf{x},\bf{{\Sigma}}} \right)$; and $\sim$ stands for ``distributed as". The symbol $j$ denotes the imaginary unit of complex numbers, with ${j^2} =  - 1$. For a set $\cal S$, $\left| {\cal S} \right|$ denotes its cardinality. The notations $\succeq$,  $\succ$ and $\preceq$ denote the componentwise inequality. ${\mathbb E}\left({\cdot}\right)$ represents the statistical expectation, and ${\cal O}\left(\cdot  \right)$ represents the standard big-O notation.

\section{System Model and Delay-Doppler Alignment Modulation}\label{sectionSystemModel}
\subsection{System Model}
 \begin{figure}[!t]
 \centering
 \centerline{\includegraphics[width=3.5in,height=2.1in]{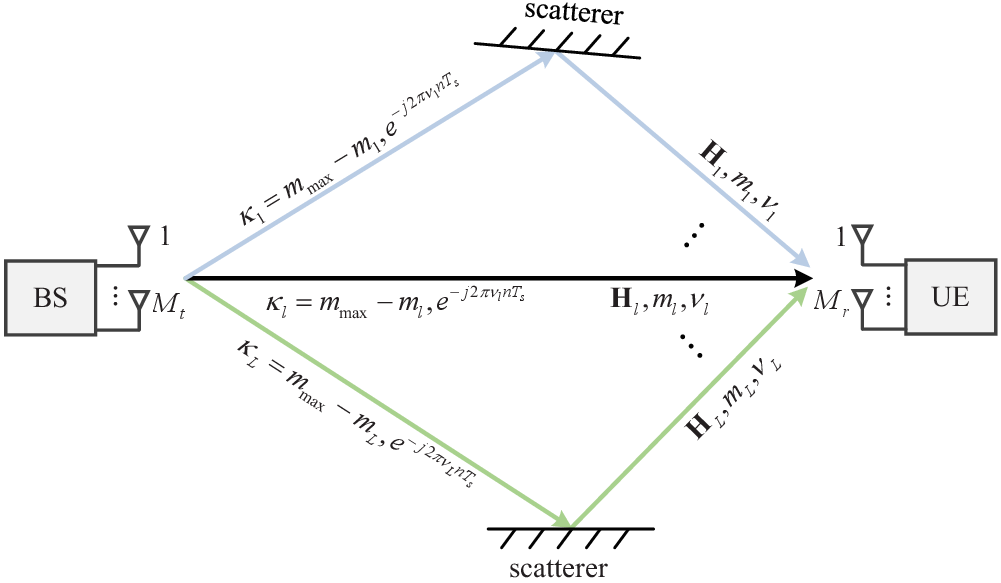}}
 \caption{A mmWave massive MIMO communication system in time-frequency doubly selective channel with delay-Doppler alignment modulation.}
 \label{systemModel}
 \end{figure}
 As shown in Fig.~\ref{systemModel}, we consider a mmWave massive MIMO communication system, where the BS and UE are equipped with $M_t \gg 1 $ and $M_r \ge 1$ antennas, respectively. The time-delay domain representation of the doubly selective channel is
 \begin{equation}\label{channelImpulseResponse}
 {\bf{H}}\left[ {n,m} \right] = \sum\limits_{l = 1}^L {{{\bf{H}}_l}{e^{j2\pi {{\nu _l}}n{T_s}}}\delta \left[ {m - {m_l}} \right]},
 \end{equation}
 where $n$ and $m$ denote the sample time and delay, respectively, $L$ is the number of multi-paths, ${\bf{H}}_l \in {{\mathbb{C}}^{M_r \times M_t}}$ denotes the channel matrix for the $l$th multi-path, ${{\nu _l}}$ and ${m_l}$ denote its Doppler frequency and delay, respectively, and $T_s = 1/B$ is the symbol duration, with $B$ representing the system bandwidth. Note that we focus on the time duration over which the number of multi-paths, delays and Doppler frequencies are approximately unchanged. Let ${m_{\min }} \triangleq \mathop {\min }\limits_{1 \le l \le L} {m_l}$ and ${m_{\max }} \triangleq \mathop {\max }\limits_{1 \le l \le L} {m_l}$ be the minimum and maximum delay over all the $L$ multi-paths, respectively. Then the channel delay spread is defined as ${m_{\rm{span}}} = {m_{\max }} - {m_{\min }}$. To study the fundamental performance limit of DDAM, perfect CSI is assumed to be available at the BS and the UE. Note that the developing trends of extremely large-scale MIMO (XL-MIMO) \cite{bjornson2019massive,lu2022communicating} and ISAC for  sixth-generation (6G) render it more feasible to obtain the CSI of the individual path information, such as AoA/AoD, delay, and Doppler frequency \cite{zhang2021enabling}. A preliminary attempt of CSI estimation for DAM is pursued in our parallel work \cite{ding2022channel}, and its extension to the considered DDAM system is important to investigate in the future.

 Let ${\bf{x}}\left[ n \right] \in {{\mathbb{C}}^{M_t \times 1}}$ denote the transmitted signal by the $M_t$ antennas. The received signal at the UE is given by
 \begin{equation}\label{generalReceivedSignal}
 \begin{aligned}
 {\bf{r}}\left[ n \right] &= {\bf{H}}\left[ {n,m} \right] \circledast {\bf{x}}\left[ n \right] + {\bf{z}}\left[ n \right]\\
 &= \sum\limits_{l = 1}^L {{{\bf{H}}_l}{e^{j2\pi {{\nu _l}}n{T_s}}}{\bf{x}}\left[ {n - {m_l}} \right]}  + {\bf{z}}\left[ n \right],
 \end{aligned}
 \end{equation}
 where ${\bf{z}}\left[ n \right] \sim {\cal CN} \left( {{\bf{0}},{\sigma ^2}{{\bf{I}}_{M_r}}} \right)$ is the additive white Gaussian noise (AWGN).

 \begin{figure*}[!t]
  \centering
  \centerline{\includegraphics[width=5.5in,height=2.0in]{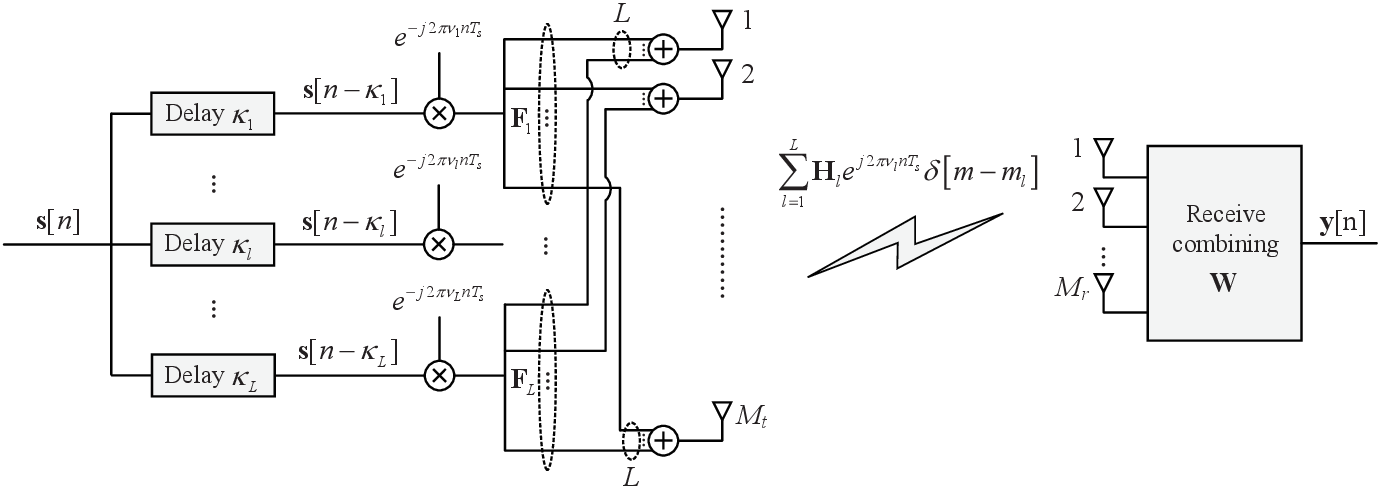}}
  \caption{Transceiver architecture of MIMO delay-Doppler alignment modulation.}
  \label{DDAMBlockDiagram}
  \end{figure*}
 Denote by ${\bf{s}}\left[ n \right] \in {{\mathbb{C}}^{{N_s} \times 1}}$ the independent and identically distributed (i.i.d.) information-bearing symbol vectors, with ${\mathbb{E}}[{\bf{s}}[n]{{\bf{s}}^H}[n]] = {{\bf{I}}_{{N_s}}}$, where ${N_s} \le \min \left\{ {{M_t},{M_r}} \right\} = {M_r}$ is the number of data streams. In this paper, we propose the DDAM transmission, which is an extension of the recently proposed DAM technique in \cite{lu2022delay} to time-frequency doubly selective MIMO channels. The transmitted signal by the BS with DDAM is
 \begin{equation}\label{DDAMTransmitSignal}
 {\bf{x}}\left[ n \right] = \sum\limits_{l = 1}^L {{{\bf{F}}_l}{\bf{s}}\left[ {n - {\kappa _l}} \right]{e^{ - j2\pi {{\nu _l}}n{T_s}}}},
 \end{equation}
 where ${{{\bf{F}}_l}} \in {{\mathbb{C}}^{{M_t} \times {N_s}}}$ denotes the path-based transmit precoding matrix associated with multi-path $l$, $\kappa _l$ is the deliberately introduced delay for the symbol vector sequence ${\bf{s}}\left[ n \right]$, with ${\kappa _l} \ne {\kappa _{l'}}$, $\forall l \ne l'$, and ${e^{ - j2\pi {{\nu _l}}n{T_s}}}$ is the introduced path-based phase pre-rotation to compensate for the Doppler frequency of multi-path $l$. The transceiver architecture for MIMO DDAM is illustrated in Fig.~\ref{DDAMBlockDiagram}. The transmit power of the BS is
 \begin{equation}\label{transmitPowerConstraint}
 \begin{aligned}
 {\mathbb{E}}\left[ {{{\left\| {{\bf{x}}\left[ n \right]} \right\|}^2}} \right] &={\mathbb{E}} \left[ {{\rm{Tr}}\left( {{\bf{x}}\left[ n \right]{{\bf{x}}^H}\left[ n \right]} \right)} \right]=  \sum\limits_{l = 1}^L {{\rm{Tr}}\left( {{{\bf{F}}_l}{\bf{F}}_l^H} \right)} \\
  &= \sum\limits_{l = 1}^L {\left\| {{{\bf{F}}_l}} \right\|_F^2}  \le P,
 \end{aligned}
 \end{equation}
 where $P$ is the available transmit power. Note that the second equality in \eqref{transmitPowerConstraint} holds since ${\bf{s}}\left[ n \right]$ are independent across different $n$ and ${\kappa _l} \ne {\kappa _{l'}}$, $\forall l\neq l'$. Denote by ${\bf{W}} \in {{\mathbb{C}}^{{M_r} \times {N_s}}}$ the receive combining matrix.{\footnote[1]{Note that per-path-based delay post-compensation and receive combining can also be applied at the UE side, but it is not considered here for simplicity.}} By substituting \eqref{DDAMTransmitSignal} into \eqref{generalReceivedSignal}, the resulting signal for DDAM after the receive combining is
 \begin{equation}\label{DAMReceivedSignal}
 \begin{aligned}
 {\bf{y}}\left[ n \right] &= {{\bf{W}}^H}{\bf{r}}\left[ n \right] = \sum\limits_{l = 1}^L {{{\bf{W}}^H}{{\bf{H}}_l}{{\bf{F}}_l}{\bf{s}}\left[ {n - {\kappa _l} - {m_l}} \right]}  + \\
 &\sum\limits_{l = 1}^L {\sum\limits_{l' \ne l}^L {{{\bf{W}}^H}{{\bf{H}}_l}{{\bf{F}}_{l'}}{\bf{s}}\left[ {n - {\kappa _{l'}} - {m_l}} \right]} } {e^{j2\pi \Delta {\nu_{l,l'}}n{T_s}}} + {\bf{\tilde z}}\left[ n \right],
 \end{aligned}
 \end{equation}
 where ${\bf{\tilde z}}\left[ n \right] \triangleq {{\bf{W}}^H}{\bf{z}}\left[ n \right]$, and $\Delta {\nu _{l,l'}} \triangleq {{\nu _l}} - {\nu _{l'}}$ denotes the \emph{Doppler difference} between multi-path $l'$ and $l$. By letting ${\kappa _l} = {m_{\max }} - {m_l} \ge 0$, $\forall l$, we have
 \begin{equation}\label{DDAMReceivedSignal}
 \begin{aligned}
 &{\bf{y}}\left[ n \right] = \underbrace {\left( {\sum\limits_{l = 1}^L {{{\bf{W}}^H}{{\bf{H}}_l}{{\bf{F}}_l}} } \right){\bf{s}}\left[ {n - {m_{\max }}} \right]}_{{\rm{desired\ signal\ with\ time-invariant\ channel}}} +  \\
 &\underbrace {\sum\limits_{l = 1}^L {\sum\limits_{l' \ne l}^L {{{\bf{W}}^H}{{\bf{H}}_l}{{\bf{F}}_{l'}}{\bf{s}}\left[ {n - {m_{\max }} + \Delta {m_{l,l'}} } \right]} } {e^{j2\pi \Delta {\nu _{l,l'}}n{T_s}}}}_{{\rm{ISI\ with\ time-variant\ channels}}} + {\bf{\tilde z}}\left[ n \right],
 \end{aligned}
 \end{equation}
 where $\Delta {m_{l,l'}} \triangleq {m_{l'}} - {m_l}$ denotes the \emph{delay difference} between path $l'$ and $l$. It is observed that by locking to the maximum delay $m_{\max}$, the first term in \eqref{DDAMReceivedSignal} is the desired signal, whose channel is contributed by all the $L$ multi-path components and is time-invariant, while the second term is the time-variant ISI due to the Doppler effect. Fortunately, the ISI can be eliminated by designing the path-based transmit precoding matrices $\{ {{\bf{F}}_l}\} _{l = 1}^L$ and the receive combining matrix $\bf{W}$ so that the following ZF constraints are satisfied
 \begin{subequations}\label{ZFDDAMCondition}
 \begin{align}
 & {{\bf{W}}^H}{{\bf{H}}_{l}}{{\bf{F}}_{l'}} = {{\bf{0}}_{{N_s} \times {N_s}}},\ \forall l \ne l', \label{ZFDDAMCondition:sub1}\\
 &{\rm{rank}}\left( {{{\bf{F}}_l}} \right) = {N_s},\ \forall l,\label{ZFDDAMCondition1:sub2}\\
 &{\rm{rank}}\left( {\bf{W}} \right) = {N_s}.\label{ZFDDAMCondition:sub3}
 \end{align}
 \end{subequations}
 In this case, the signal in \eqref{DDAMReceivedSignal} reduces to
 \begin{equation}\label{DDAMReceivedSignalEquivalent}
 {\bf{y}}\left[ n \right] = \left(\sum\limits_{l = 1}^L {{{\bf{W}}^H}{{\bf{H}}_l}{{\bf{F}}_l}} \right){\bf{s}}\left[ {n - {m_{\max }}} \right] + {\bf{\tilde z}}\left[ n \right].
 \end{equation}

 It is observed that the resulting signal in \eqref{DDAMReceivedSignalEquivalent} is simply the symbol vector sequence ${\bf{s}}\left[ n \right]$ delayed by one single delay $m_{\max}$, which is free from the detrimental ISI. Furthermore, the resulting channel gain is contributed by all the $L$ multi-path channel components. Besides, thanks to the Doppler pre-compensation in \eqref{DDAMTransmitSignal}, the Doppler effect has been eliminated, and the channel is no longer time-variant. In other words, with the proposed DDAM in \eqref{DDAMTransmitSignal} and the ZF design in \eqref{ZFDDAMCondition}, the original time-variant frequency-selective channel has been transformed to the simple time-invariant ISI-free channel.

 \subsection{ZF Conditions}
 Note that the ZF conditions in \eqref{ZFDDAMCondition} are similar to those in the extensively studied interference alignment scheme \cite{cadambe2008interference,gomadam2008approaching,yetis2010feasibility,razaviyayn2012degrees,gomadam2011distributed,sridharan2015degrees}, while a subtle difference lies in that DDAM is on the per-path basis, and interference alignment is on the per-user basis. In general, the feasibility of ZF conditions to achieve interference alignment remains an open problem \cite{sridharan2015degrees}, and some useful results have been obtained in e.g., \cite{yetis2010feasibility,razaviyayn2012degrees,gomadam2011distributed}. By following similar procedures in \cite{yetis2010feasibility,razaviyayn2012degrees}, we have the following results.
 \begin{figure}[!t]
  \centering
  \subfigure[$M_r = N_s = 1$.]{
    \includegraphics[width=1.55in,height=1.5in]{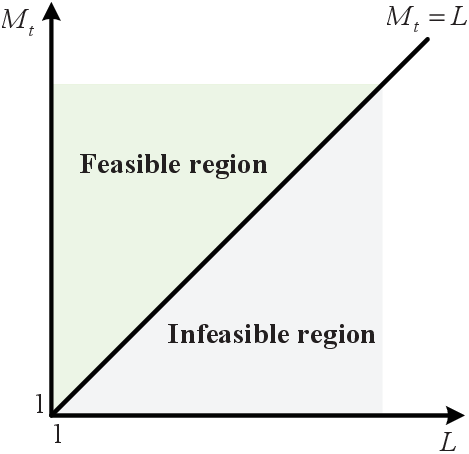}
  }
  \subfigure[$M_r = N_s = 2$.]{
    \includegraphics[width=1.55in,height=1.6in]{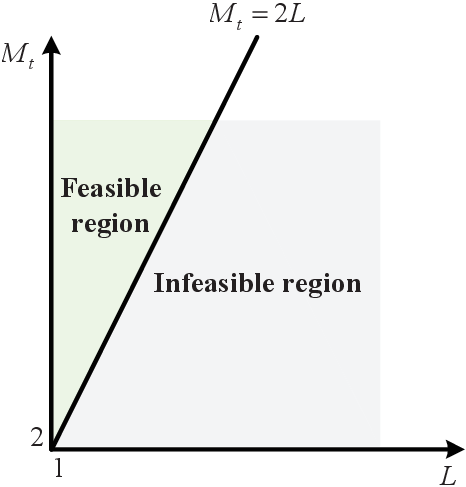}
  }
  \subfigure[$M_r = N_s = 4$.]{
    \includegraphics[width=1.55in,height=1.75in]{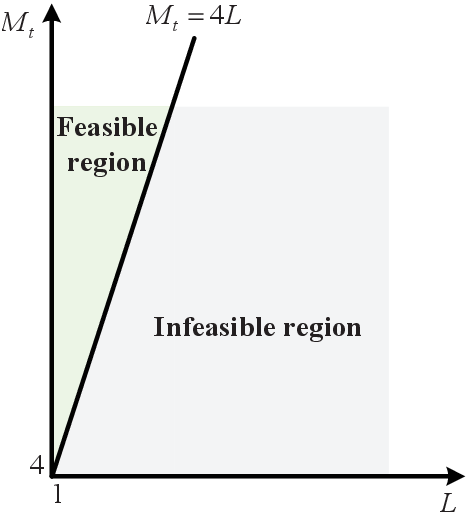}
  }
  \subfigure[$M_r = 4$, $N_s = 2$. ]{
    \includegraphics[width=1.55in,height=1.8in]{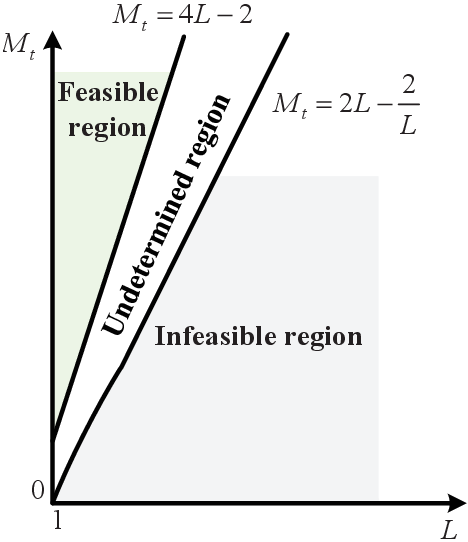}
  }
  \caption{Illustration of feasible and infeasible regions to achieve the ZF conditions \eqref{ZFDDAMCondition}.}
  \label{feasibleConditionIllustration}
 \end{figure}

 \begin{proposition} \label{ZFInfeasibilityProposition}
 One necessary condition for the ZF conditions in \eqref{ZFDDAMCondition} to be feasible is
 \begin{equation}\label{ZFNecessaryCondition}
 L{M_t} + {M_r} \ge \left( {{L^2} + 1} \right){N_s}.
 \end{equation}
 \end{proposition}

 \begin{IEEEproof}
 Please refer to Appendix~\ref{ProofZFInfeasibilityProposition}.
 \end{IEEEproof}

 \begin{proposition} \label{ZFSufficientConditionProposition}
 One sufficient condition for the ZF conditions in \eqref{ZFDDAMCondition} to be feasible is
 \begin{equation}\label{ZFSufficientCondition}
 {M_t} \ge \left( {L - 1} \right){M_r} + {N_s}.
 \end{equation}
 \end{proposition}
 \begin{IEEEproof}
 Please refer to Appendix~\ref{ProofZFSufficientConditionProposition}.
 \end{IEEEproof}

 \begin{proposition} \label{MISOZFConditionProposition}
 If ${N_s} = {M_r}$, the ZF conditions in \eqref{ZFDDAMCondition} are feasible if and only if ${M_t} \ge L{N_s}$.
 \end{proposition}
 \begin{IEEEproof}
 Proposition~\ref{MISOZFConditionProposition} can be shown by noting that when ${N_s} = {M_r} $, both \eqref{ZFNecessaryCondition} and \eqref{ZFSufficientCondition} reduce to ${M_t} \ge L{N_s}$.
 \end{IEEEproof}

 Fig. \ref{feasibleConditionIllustration} shows the feasible and infeasible regions of $M_t$ versus $L$ specified by the necessary condition \eqref{ZFNecessaryCondition} and the sufficient condition \eqref{ZFSufficientCondition} for different values of $M_r$ and $N_s$. It is observed from Figs. \ref{feasibleConditionIllustration}(a)-(c) that when ${N_s} = {M_r}$, the sufficient and necessary condition to achieve the ZF requirement in \eqref{ZFDDAMCondition} is ${M_t} \ge L{N_s}$, as specified in Proposition~\ref{MISOZFConditionProposition}. This makes intuitive sense since when $M_t \ge L{N_s}$, all the data streams over the multi-paths can be separated at the BS, thus guaranteeing the ZF conditions. It is also observed from Fig. \ref{feasibleConditionIllustration}(d) that when ${N_s} < {M_r}$, there still exists undetermined region.

 \section{Asymptotic Analysis For DDAM}\label{sectionAsymptoticAnalysis}
 In this subsection, we provide the asymptotic analysis by assuming that the number of BS antennas is much larger than the number of channel paths, i.e., ${M_t} \gg L$. For ease of illustration, we assume that single data stream is transmitted, i.e., $N_s = 1$. In this case, the transmitted signal of DDAM in \eqref{DDAMTransmitSignal} reduces to
 \begin{equation}\label{DDAMTransmitSignalSingleStream}
 {\bf{x}}\left[ n \right] = \sum\limits_{l = 1}^L {{{\bf{f}}_l}s\left[ {n - {\kappa _l}} \right]{e^{ - j2\pi {{\nu _l}}n{T_s}}}}.
 \end{equation}
 With the receive combining vector ${\bf{w}} \in {{\mathbb C}^{{M_r} \times 1}}$, where $\left\| {\bf{w}} \right\| = 1$, the signal in \eqref{DDAMReceivedSignal} reduces to
 \begin{equation}\label{DDAMReceivedSignalSingleStream}
 \begin{aligned}
 &y\left[ n \right] = \underbrace {\left( {\sum\limits_{l = 1}^L {{{\bf{w}}^H}{{\bf{H}}_l}{{\bf{f}}_l}} } \right)s\left[ {n - {m_{\max }}} \right]}_{{\rm{desired\ signal\ with\ time-invariant\ channel}}} +  \\
 &\underbrace {\sum\limits_{l = 1}^L {\sum\limits_{l' \ne l}^L {{{\bf{w}}^H}{{\bf{H}}_l}{{\bf{f}}_{l'}}s\left[ {n - {m_{\max }} + \Delta {m_{l,l'}}} \right]} } {e^{j2\pi \Delta {\nu _{l,l'}}n{T_s}}}}_{{\rm{ISI\ with\ time-variant\ channels}}} + \tilde z\left[ n \right],
 \end{aligned}
 \end{equation}
 where $\tilde z\left[ n \right] \triangleq {{\bf{w}}^H}{\bf{z}}\left[ n \right]$.

 Denote by ${\phi _l}$ and ${\varphi _l}$ the AoA and AoD of the $l$th multi-path, respectively. The channel matrix of multi-path $l$ can be expressed as ${{\bf{H}}_l} = {\alpha _l}{{\bf{a}}_R}\left( {{\phi _l}} \right){\bf{a}}_T^H\left( {{\varphi _l}} \right)$, where ${\alpha _l}$ is the complex-valued channel coefficient, and ${{\bf{a}}_R}\left( {{\phi _l}} \right) \in {{\mathbb C}^{{M_r} \times 1}}$ and ${\bf{a}}_T\left( {{\varphi _l}} \right) \in {{\mathbb C}^{{M_t} \times 1}}$ represent the receive and transmit array response vectors, respectively. Note that for the standard far-field uniform plane wave (UPW) model, we have ${\left\| {{{\bf{a}}_T}\left( {{\varphi _l}} \right)} \right\|^2} = {M_t}$ and ${\left\| {{{\bf{a}}_R}\left( {{\phi _l}} \right)} \right\|^2} = {M_r}$. Let ${{{\bf{\bar a}}}_T}\left( {{\varphi _l}} \right) \triangleq \alpha _l^*{{\bf{a}}_T}\left( {{\varphi _l}} \right)$, according to the results in \cite{lu2023Manipulating}, as long as the $L$ multi-paths are associated with different AoDs, $\{ {{{\bf{\bar a}}}_T}\left( {{\varphi _l}} \right)\} _{l = 1}^L$ tend to be asymptotically orthogonal when ${M_t} \gg L$, i.e.,
 \begin{equation}\label{asymptoticalProperty1}
 \frac{{{\bf{\bar a}}_T^H\left( {{\varphi _l}} \right){{{\bf{\bar a}}}_T}\left( {{\varphi _{l'}}} \right)}}{{\left\| {{{{\bf{\bar a}}}_T}\left( {{\varphi _l}} \right)} \right\|\left\| {{{{\bf{\bar a}}}_T}\left( {{\varphi _{l'}}} \right)} \right\|}} \to 0,\ \forall l \ne l'.
 \end{equation}
 Furthermore, since ${{\bf{H}}_l} = {\alpha _l}{{\bf{a}}_R}\left( {{\phi _l}} \right){\bf{a}}_T^H\left( {{\varphi _l}} \right)$, we have
 \begin{equation}\label{asymptoticalProperty2}
 \frac{{{{\bf{H}}_l}{\bf{H}}_{l'}^H}}{{\left\| {{{{\bf{\bar a}}}_T}\left( {{\varphi _l}} \right)} \right\|\left\| {{{{\bf{\bar a}}}_T}\left( {{\varphi _{l'}}} \right)} \right\|}} \to {\bf{0}},\ \forall l \ne l'.
 \end{equation}

 By applying the simple path-based MRT beamforming, each transmit beamforming vector ${{\bf{f}}_l}$ is given by
 \begin{equation}\label{pathbasedMRTBeamforming}
 {{\bf{f}}_l} = \sqrt {{p_l}} \frac{{{{{\bf{\bar a}}}_T}\left( {{\varphi _l}} \right)}}{{\left\| {{{{\bf{\bar a}}}_T}\left( {{\varphi _l}} \right)} \right\|}} = \sqrt {{p_l}} \frac{{\alpha _l^*{{\bf{a}}_T}\left( {{\varphi _l}} \right)}}{{\left| {{\alpha _l}} \right|\sqrt {{M_t}} }},\ \forall l,
 \end{equation}
 where $p_l$ denotes the power allocation for the $l$th multi-path. By substituting \eqref{pathbasedMRTBeamforming} into \eqref{DDAMReceivedSignalSingleStream} and scaling the resulting signal ${\bf{y}}\left[ n \right]$ by $\xi  \triangleq 1/\sum\nolimits_{i = 1}^L {\left\| {{\bf{\bar a}}\left( {{\varphi _i}} \right)} \right\|} $, we have \eqref{ReceivedSignalMRT} shown at the top of the next page.
 \newcounter{mytempeqncnt1}
 \begin{figure*}
 \normalsize
 \setcounter{mytempeqncnt1}{\value{equation}}
 \begin{equation}\label{ReceivedSignalMRT}
 \begin{aligned}
 \xi y\left[ n \right] &= \xi \sqrt {{M_t}} \left({{\bf{w}}^H} {\sum\limits_{l = 1}^L {\sqrt {{p_l}} \left| {{\alpha _l}} \right|{{\bf{a}}_R}\left( {{\phi _l}} \right)} } \right)s\left[ {n - {m_{\max }}} \right] +  \\
 &\ \ \ \ \ \sum\limits_{l = 1}^L {\sum\limits_{l' \ne l}^L {{{\bf{w}}^H} \sqrt {{p_{l'}}} {{\bf{a}}_R}\left( {{\phi _l}} \right)\frac{{{\bf{\bar a}}_T^H\left( {{\varphi _l}} \right){{{\bf{\bar a}}}_T}\left( {{\varphi _{l'}}} \right)}}{{\left( {\sum\nolimits_{i = 1}^L {\left\| {{\bf{\bar a}}\left( {{\varphi _i}} \right)} \right\|} } \right)\left\| {{{{\bf{\bar a}}}_T}\left( {{\varphi _{l'}}} \right)} \right\|}}s\left[ {n - {m_{\max }} + \Delta {m_{l,l'}}} \right]} } {e^{j2\pi \Delta {\nu _{l,l'}}n{T_s}}} + \xi \tilde z\left[ n \right].
 \end{aligned}
 \end{equation}
 \hrulefill
 \end{figure*}
 In particular, thanks to the asymptotically orthogonal property brought by large antenna arrays, we have
 \begin{equation}\label{asymptoticalProperty3}
 \frac{{{\bf{\bar a}}_T^H\left( {{\varphi _l}} \right){{{\bf{\bar a}}}_T}\left( {{\varphi _{l'}}} \right)}}{{\left( {\sum\nolimits_{i = 1}^L {\left\| {{\bf{\bar a}}\left( {{\varphi _i}} \right)} \right\|} } \right)\left\| {{{{\bf{\bar a}}}_T}\left( {{\varphi _{l'}}} \right)} \right\|}} \le \frac{{{\bf{\bar a}}_T^H\left( {{\varphi _l}} \right){{{\bf{\bar a}}}_T}\left( {{\varphi _{l'}}} \right)}}{{\left\| {{\bf{\bar a}}\left( {{\varphi _l}} \right)} \right\|\left\| {{{{\bf{\bar a}}}_T}\left( {{\varphi _{l'}}} \right)} \right\|}} \to 0.
 \end{equation}
 With \eqref{asymptoticalProperty3}, and by dividing \eqref{ReceivedSignalMRT} with $\xi$, the resulting signal in \eqref{DDAMReceivedSignalSingleStream} reduces to
 \begin{equation}\label{ReceivedSignalMRT2}
 y\left[ n \right] \to \sqrt {{M_t}} \left({{\bf{w}}^H} {\sum\limits_{l = 1}^L {\sqrt {{p_l}} \left| {{\alpha _l}} \right|{{\bf{a}}_R}\left( {{\phi _l}} \right)} } \right)s\left[ {n - {m_{\max }}} \right]  + \tilde z\left[ n \right].
 \end{equation}
 As can be seen in \eqref{ReceivedSignalMRT2}, when ${M_t} \gg L$, the original time-variant frequency-selective channel has been transformed into the time-invariant ISI-free channel with the simple delay-Doppler compensation and path-based MRT beamforming, without resorting to the sophisticated channel equalization or multi-carrier transmission. The resulting signal-to-noise ratio (SNR) is
 \begin{equation}\label{ReceivedSNRMRT}
 \gamma  = \frac{{{M_t}{{\left| {{{\bf{w}}^H}\sum\limits_{l = 1}^L {\sqrt {{p_l}} \left| {{\alpha _l}} \right|{{\bf{a}}_R}\left( {{\phi _l}} \right)} } \right|}^2}}}{{{\sigma ^2}}}.
 \end{equation}
 For any given power allocation $\left\{ {{p_l}} \right\}_{l = 1}^L$, by applying Cauchy-Schwarz inequality, the optimal combining vector ${{\bf w}^\star}$ is
 \begin{equation}\label{receiveBeamformingMRT}
 {{\bf{w}}^ \star } = \frac{{\sum\limits_{l = 1}^L {\sqrt {{p_l}} \left| {{\alpha _l}} \right|{{\bf{a}}_R}\left( {{\phi _l}} \right)} }}{{\left\| {\sum\limits_{l = 1}^L {\sqrt {{p_l}} \left| {{\alpha _l}} \right|{{\bf{a}}_R}\left( {{\phi _l}} \right)} } \right\|}},
 \end{equation}
 and the resulting SNR can be expressed as
 \begin{equation}\label{ReceivedSNRMRT2}
 \gamma  = \frac{{{M_t}{{\left\| {\sum\limits_{l = 1}^L {\sqrt {{p_l}} \left| {{\alpha _l}} \right|{{\bf{a}}_R}\left( {{\phi _l}} \right)} } \right\|}^2}}}{{{\sigma ^2}}}.
 \end{equation}
 By applying the triangle inequality, we have the upper bound of \eqref{ReceivedSNRMRT2}, given by
 \begin{equation}\label{ReceivedSNRMRT2UpperBound}
 \gamma  \le \frac{{{M_t}{{\left( {\sum\limits_{l = 1}^L {\sqrt {{p_l}} \left| {{\alpha _l}} \right|\left\| {{{\bf{a}}_R}\left( {{\phi _l}} \right)} \right\|} } \right)}^2}}}{{{\sigma ^2}}} = \frac{{{M_t}{M_r}{{\left( {\sum\limits_{l = 1}^L {\sqrt {{p_l}} \left| {{\alpha _l}} \right|} } \right)}^2}}}{{{\sigma ^2}}}.
 \end{equation}

 \begin{proposition} \label{ReceivedSNRMRT2OptimalUpperBound}
 The maximal SNR upper bound in \eqref{ReceivedSNRMRT2UpperBound} with the optimal power allocation is
 \begin{equation}\label{ReceivedSNRMRT2OptimalSNR}
 {\gamma ^ \star } = \bar P{M_t}{M_r}\sum\nolimits_{l = 1}^L {{{\left| {{\alpha _l}} \right|}^2}},
 \end{equation}
 where ${\bar P} \triangleq P/{\sigma ^2}$.
 \end{proposition}

 \begin{IEEEproof}
 Please refer to Appendix~\ref{ProofReceivedSNRMRT2OptimalUpperBound}.
 \end{IEEEproof}

 As a comparison, the single-carrier benchmarking scheme termed {\it strongest-path beamforming} is considered, which only uses the single dominant path. Specifically, the transmit signal of the strongest path beamforming scheme is
 \begin{equation}\label{SPMTransmitSignal}
 {\bf{x}}\left[ n \right] = {\bf{f}}s\left[ n \right].
 \end{equation}
 Similarly, by applying the simple MRT beamforming ${\bf{f}} = \sqrt P {{\bf{\bar a}}_T}\left( {{\varphi _{{l_{\max }}}}} \right)/\left\| {{{{\bf{\bar a}}}_T}\left( {{\varphi _{{l_{\max }}}}} \right)} \right\|$, with ${l_{\max }} \triangleq \arg \mathop {\max }\limits_{1 \le l \le L} {\left| {{\alpha _l}} \right|^2}$, the resulting signal after the receive combining is
 \begin{equation}\label{SPMResultingSignal}
 y\left[ n \right] {\rm =} \sqrt P \sum\limits_{l = 1}^L {{{\bf{w}}^H}{{\bf{H}}_l}\frac{{{{\bf{\bar a}}_T}({\varphi _{{l_{\max }}}})}}{{\left\| {{{\bf{\bar a}}_T}({\varphi _{{l_{\max }}}})} \right\|}}s\left[ {n - {m_l}} \right]{e^{j2\pi {\nu _l}n{T_s}}}}  + \tilde z\left[ n \right].
 \end{equation}
 With the asymptotically orthogonal property and the optimal combining vector ${{\bf{w}}^{\rm{ \star }}} = {{\bf{a}}_R}\left( {{\phi _{{l_{\max }}}}} \right)/\left\| {{{\bf{a}}_R}\left( {{\phi _{{l_{\max }}}}} \right)} \right\|$, the resulting signal in \eqref{SPMResultingSignal} reduces to
 \begin{equation}\label{SPMReducedResultingSignal}
 y\left[ n \right] \to \sqrt P \sqrt {{M_t}{M_r}} \left| {{\alpha _{{l_{\max }}}}} \right|s[n - {m_{{l_{\max }}}}]{e^{j2\pi {\nu _{{l_{\max }}}}n{T_s}}} + \tilde z\left[ n \right].
 \end{equation}
 The optimal SNR of the strongest path MRT beamforming is
 \begin{equation}\label{SPBOptimalSNR}
 {\gamma ^ \star } = \bar P{M_t}{M_r}{\left| {{\alpha _{{l_{\max }}}}} \right|^2}.
 \end{equation}
 It is observed from \eqref{ReceivedSNRMRT2OptimalSNR} and \eqref{SPBOptimalSNR} that DDAM outperforms the strongest path beamforming scheme via benefiting from all the multi-path components. Note that the gain between DDAM and the strongest path beamforming scheme will become even more significant when there are several comparable dominating paths. For example, besides the inherent multi-path components in the environment, a few strong virtual line-of-sight (LoS) components can be created by deploying the intelligent reflecting surfaces (IRSs) \cite{lu2023delayIRS} or metal reflectors \cite{yu2023wireless}, and a significant performance gain is expected in such scenarios. In the following two sections, we study the DDAM design for the general case without assuming ${M_t} \gg L$.

\section{Path-Based ZF Precoding Towards Time-Invariant ISI-free DDAM Communication}\label{sectionPathBasedZFPrecoding}
 In this section, we study the path-based ZF precoding and receive combining optimization for DDAM communication, so that the ISI in \eqref{DDAMReceivedSignal} is eliminated and the original time-variant frequency-selective channel is transformed to time-invariant ISI-free channel. As stated in Proposition \ref{ZFSufficientConditionProposition}, one sufficient (but not necessary) condition for the ZF requirement is given in \eqref{ZFSufficientCondition}. It is worth mentioning that the assumptions of ${M_t} \gg L$ and ${N_s}=1$ used in Section \ref{sectionAsymptoticAnalysis} are no longer needed here.

 By designing the path-based precoding matrices $\{ {{\bf{F}}_l}\} _{l = 1}^L$ as \eqref{pathBasedZFBeamforming} in Appendix~\ref{ProofZFSufficientConditionProposition}, ${{\bf{F}}_{l}}$ can be expressed as
 ${{\bf{F}}_{l}} = {\bf{\tilde H}}_{l}^ \bot {{\bf{X}}_{l}}$, where ${\bf{\tilde H}}_{l}^ \bot  \in {{\mathbb C}^{{M_t} \times {r_{l}}}}$ is an orthonormal basis for the orthogonal complement of ${{{\bf{\tilde H}}}_{l}}$, with ${{{\bf{\tilde H}}}_{l}} \in {{\mathbb C}^{{M_t} \times \left( {L - 1} \right){M_r}}} = \left[ {{\bf{H}}_1^H, \cdots ,{\bf{H}}_{l - 1}^H,{\bf{H}}_{l + 1}^H, \cdots ,{\bf{H}}_L^H} \right]$ defined below \eqref{pathBasedZFBeamforming} and ${r_{l}} = {\rm{rank}}( {{\bf{\tilde H}}_{l}^ \bot }) = {M_t} - \left( {L - 1} \right){M_r}$, and ${{\bf{X}}_{l}} \in {{\mathbb C}^{{r_{l}} \times {N_s}}}$ is the new transmit precoding matrix to be designed. By substituting ${{\bf{F}}_{l}} = {\bf{\tilde H}}_{l}^ \bot {{\bf{X}}_{l}}$ into \eqref{DDAMReceivedSignal}, the resulting signal reduces to
 \begin{equation}\label{DDAMReceivedSignalZF}
 \begin{aligned}
 {\bf{y}}\left[ n \right] &= \left( {\sum\limits_{l = 1}^L {{{\bf{W}}^H}{{\bf{H}}_l}{\bf{\tilde H}}_l^ \bot {{\bf{X}}_l}} } \right){\bf{s}}\left[ {n - {m_{\max }}} \right] + {\bf{\tilde z}}\left[ n \right]\\
 & = {{\bf{W}}^H}{\bf{{\tilde H}{\tilde X}}}{\bf s}\left[ {n - {m_{\max }}} \right] + {\bf{\tilde z}}\left[ n \right],
 \end{aligned}
 \end{equation}
 where ${\bf{\tilde H}} \in {{\mathbb C}^{{M_r} \times {r_{\rm{sum}}} }} \triangleq [ {{{\bf{H}}_1}{\bf{\tilde  H}}_1^ \bot , \cdots ,{{\bf{H}}_L}{\bf{\tilde  H}}_L^ \bot } ]$, ${\bf{\tilde  X}} \in {{\mathbb C}^{{r_{\rm{sum}}} \times {N_s}}} \triangleq {\left[ {{\bf{X}}_1^T, \cdots ,{\bf{X}}_L^T} \right]^T}$, with ${r_{\rm{sum}}} = \sum {{r_l}}  = L{M_t} - L\left( {L - 1} \right){M_r}$, and ${\mathbb E}\left[ {{\bf{\tilde z}}\left[ n \right]{{{\bf{\tilde z}}}^H}\left[ n \right]} \right] = {\sigma ^2}{{\bf{W}}^H}{\bf{W}}$. It is observed from \eqref{DDAMReceivedSignalZF} that similar to \eqref{DDAMReceivedSignalEquivalent}, the original doubly selective channel is also transformed into the time-invariant ISI-free channel. Besides, the power constraint in \eqref{transmitPowerConstraint} can be re-expressed as
 \begin{equation}\label{powerConstraintZF}
 \begin{aligned}
 \sum\limits_{l = 1}^L {\left\| {{{\bf{F}}_l}} \right\|_F^2}  &  {\rm =} {\rm{Tr}}\left( {\sum\limits_{l = 1}^L {{\bf{F}}_l^H{{\bf{F}}_l}} } \right){\rm =} {\rm{Tr}}\left( {\sum\limits_{l = 1}^L {{\bf{X}}_l^H{{\left( {{\bf{\tilde  H}}_l^ \bot } \right)}^H}{\bf{\tilde  H}}_l^ \bot {{\bf{X}}_l}} } \right)\\
 & {\rm =} {\rm{Tr}}\left( {\sum\limits_{l = 1}^L {{\bf{X}}_l^H{{\bf{X}}_l}} } \right) {\rm =} {\rm{Tr}}\left( {{{{\bf{\tilde  X}}}^H}{\bf{\tilde  X}}} \right) {\rm =} \left\| {{\bf{\tilde  X}}} \right\|_F^2,
 \end{aligned}
 \end{equation}
 where the third equality holds since ${{\bf{\tilde  H}}_l^ \bot }$ is an orthonormal basis with ${( {{\bf{\tilde  H}}_l^ \bot })^H}{\bf{\tilde  H}}_l^ \bot  = {\bf{I}}$.

 The spectral efficiency of \eqref{DDAMReceivedSignalZF} in bits/second/Hz (bps/Hz) can be maximized by jointly optimizing the transmit precoding matrix ${\bf{\tilde  X}}$ and the receive combining matrix ${\bf{W}}$, which yields the following problem
 \begin{equation}\label{originalProblemZF}
 \begin{aligned}
 \left( {\rm{P{\text -}ZF}} \right)\ \mathop {\max }\limits_{{\bf{W}},{\bf{\tilde  X}}}&\ \  {\log _2}\left| {{{\bf{I}}_{{N_s}}} + \frac{1}{{{\sigma ^2}}}{{\bf{W}}^H}{\bf{\tilde H\tilde  X}}{{{\bf{\tilde  X}}}^H}{{{\bf{\tilde  H}}}^H}{\bf{W}}{{\left( {{{\bf{W}}^H}{\bf{W}}} \right)}^{ - 1}}} \right|\\
 {\rm{s.t.}}&\ \ \left\| {{\bf{\tilde  X}}} \right\|_F^2 \le P. \nonumber
 \end{aligned}
 \end{equation}

 This corresponds to the standard MIMO capacity maximization problem, whose optimal solution can be obtained by eigendecomposition and water-filling (WF) power allocation \cite{goldsmith2005wireless}. Specifically, let the (reduced) singular value decomposition (SVD) of ${\bf{\tilde  H}}$ be expressed as ${\bf{\tilde  H}} = {\bf{\tilde R \tilde \Sigma }}{{\bf{\tilde T}}^H}$, where ${\bf{\tilde R}} \in {{\mathbb C}^{{M_r} \times {\tilde r}}}$, ${\bf{\tilde  T}} \in {{\mathbb C}^{{r_{\rm{sum}}} \times {\tilde r}}}$, and ${\bf{\tilde \Sigma }} \in {{\mathbb C}^{{\tilde r} \times {\tilde r}}}$ is a diagonal matrix of singular values. Then the optimal receive combining and transmit precoding matrices are respectively given by ${{\bf{W}} } = {\bf{\tilde R}}$ and ${\bf{\tilde X}} = {\bf{\tilde T}}{{{\bf{\tilde \Lambda }}}^{1/2}}$, where ${\bf{\tilde \Lambda }} \in {{\mathbb C}^{\tilde r \times \tilde r}}$ is the power allocation matrix, with ${\bf{\tilde \Lambda }} \triangleq {\rm{diag}}( {{{\tilde \lambda }_1}, \cdots ,{{\tilde \lambda }_{\tilde r}}} )$ \cite{goldsmith2005wireless}. As a result, (P-ZF) reduces to finding the optimal power allocation, which can be obtained by the classic WF strategy, and the details are omitted for brevity. The main complexity of path-based ZF precoding and receive combining lies in the SVD, which requires ${\cal O}\left( {{M_r}{r_{{\rm{sum}}}}\min \left( {{M_r},{r_{{\rm{sum}}}}} \right)} \right)$.

\section{Joint Path-Based Precoding and Combining Optimization for DDAM}\label{sectionJointPrecodingCombining}
 In this section, we consider the more general design by tolerating some residual ISI in \eqref{DDAMReceivedSignal}. In this case, the spectral efficiency is maximized by jointly optimizing the path-based transmit precoding matrices $\left\{ {{{\bf{F}}_l}} \right\}_{l = 1}^L$ and receive combining matrix $\bf{W}$ without imposing the ZF constraints in \eqref{ZFDDAMCondition}.

 Since the ISI in \eqref{DDAMReceivedSignal} involves time-variant channels due to Doppler effect, we first exploit the double timescales for the time-frequency channel and the state of the individual paths (e.g., AoAs/AoDs, delays and Doppler frequencies) to derive the spectral efficiency with time-variant ISI channels. Specifically, the wireless channel that is composed by multi-paths usually changes at much faster time-scales than the state of the individual paths, say on the order of milliseconds for the former versus hundreds of milliseconds for the latter \cite{xiao2022exploiting,duel2007fading,rangan2014millimeter}. Such an observation motivates the exploitation of double timescales to derive the spectral efficiency involving time-variant ISI channels. Specifically, let ${T_c} \triangleq \zeta /{\nu _{\max }}$ denote the \emph{channel coherence time}, within which the channel impulse response in \eqref{channelImpulseResponse} remains approximately unchanged, where $\zeta $ is some coefficient and ${\nu _{\max }} = \mathop {\max }\limits_{1 \le l \le L} {\nu _l}$. Moreover, let $\bar T \gg {T_c}$ denote the new timescale termed \emph{path invariant time} \cite{xiao2022exploiting}, within which the state of the multi-paths is nearly constant. With $\zeta \ll 1$, the phase rotation caused by the Doppler difference $\Delta {\nu _{l,l'}}$ in \eqref{DDAMReceivedSignal} can be negligible within each channel coherence time $T_c$. For example, for a mmWave system where the carrier frequency is $f = 28$ GHz and the total bandwidth is $B = 100$ MHz. The moving velocity of UE is $v = 180$ km/h, corresponding to the maximum Doppler frequency ${\nu _{\max }} = vf/c = 4666.7$ Hz, where $c$ is the speed of light. For illustration, the Doppler frequencies of two multi-paths are set as ${\nu_1} = 4666.7$ Hz and ${\nu _2} = 666.7$ Hz, respectively, and the Doppler difference is $\Delta {\nu_{1,2}} = 4000$ Hz. By setting $\zeta = 0.1$, the channel coherence time is ${T_c} = 0.1/{\nu_{\max}} \approx 0.0214$ ms, and the total number of signal samples within each channel coherence time is ${N_c} = B{T_c} = 2142$. Under this setup, the maximum phase rotation caused by the Doppler difference within each channel coherence time is $2\pi \Delta {\nu_{1,2}}{N_c}{T_s} = 0.17\pi$, which is relatively small. Thus, for any $n \in \left[ {1,{\bar N}} \right]$, with ${\bar N} = {\bar T}/{T_s}$, the term ${e^{j2\pi \Delta {\nu _{l,l'}}n{T_s}}}$ in \eqref{DDAMReceivedSignal} is approximated as ${e^{j2\pi \Delta {\nu _{l,l'}}n{T_s}}} \approx {e^{j2\pi \Delta {\nu _{l,l'}}\left\lceil {\frac{n}{{{N_c}}}} \right\rceil {N_c}{T_s}}}$, which remains constant within each channel coherence block, but varies across different channel coherence blocks.

 Next, we focus on the joint transmit precoding and receive combining design within one channel path invariant block. In order to derive the spectral efficiency of \eqref{DDAMReceivedSignal} with residual ISI, we need to group those interfering symbols in the form of identical delay difference, since they correspond to identical symbols \cite{zeng2018multi,lu2022delay}. Let ${\cal L} \triangleq \left\{ {l:l = 1, \cdots ,L} \right\}$ denote the set of all multi-paths, and ${{\cal L}_l} \triangleq {\cal L}\backslash l$ contains all multi-paths excluding path $l$. In particular, $\forall l \ne l'$, we have $\Delta {m_{l,l'}} \ne 0$, and $\Delta {m_{l,l'}} \in \left\{ { \pm 1, \cdots , \pm {m_{\rm{span}}}} \right\}$. Then for each delay difference $i \in \left\{ { \pm 1, \cdots , \pm {m_{\rm{span}}}} \right\}$, define the following effective channel
 \begin{equation}\label{delayDifferenceChannel}
 {\bf{G}}_{l'}\left[ i \right] \triangleq \left\{ \begin{split}
 &{{\bf{H}}_l}{e^{j2\pi \Delta {\nu _{l,l'}}\left\lceil {\frac{n}{{{N_c}}}} \right\rceil {N_c}{T_s}}},\ {\rm{if}}\ \exists l \in {{\cal L}_{l'}},\ {\rm{s.t.}}\ \Delta {m_{l,l'}} = i,\\
 &{\bf{0}}_{{M_r} \times {M_t}},\ \ \ \ \ \ \ \ \ \ \ \ \ \ \ \ \ {\rm{otherwise}}.
 \end{split} \right.
 \end{equation}
 As a result, the resulting signal in \eqref{DDAMReceivedSignal} can be equivalently written as
 \begin{equation}\label{DDAMEquivalentReceivedSignal}
 \begin{aligned}
  &{\bf{y}}\left[ n \right]= \left( {\sum\limits_{l = 1}^L {{{\bf{W}}^H}{{\bf{H}}_l}{{\bf{F}}_l}} } \right){\bf{s}}\left[ {n - {m_{\max }}} \right] +  \\
 &\sum\limits_{i =  - {m_{{\rm{span}}}},i \ne 0}^{{m_{{\rm{span}}}}} {\left( {\sum\limits_{l' = 1}^L {{{\bf{W}}^H}{{\bf{G}}_{l'}}\left[ i \right]{{\bf{F}}_{l'}}} } \right)}{\bf{s}}\left[ {n - {m_{\max }} + i} \right] + {\bf{\tilde z}}\left[ n \right].
 \end{aligned}
 \end{equation}

 Let ${\bf{\bar H}} = \left[ {{{\bf{H}}_1}, \cdots ,{{\bf{H}}_L}} \right] \in {{\mathbb{C}}^{{M_r} \times {LM_t}}} $, ${\bf{\bar F}} = \left[ {{\bf{F}}_1^T, \cdots ,} \right.$ $ {\left. {{\bf{F}}_L^T} \right]^T} \in {{\mathbb{C}}^{{LM_t} \times {N_s}}} $, and ${\bf{\bar G}}\left[ i \right] = \left[ { {{\bf{G}}_1}\left[ i \right], \cdots , {{\bf{G}}_L}\left[ i \right]} \right] \in {{\mathbb{C}}^{ {M_r}\times {LM_t}}}$,  \eqref{DDAMEquivalentReceivedSignal} is compactly written as
 \begin{equation}\label{DDAMEquivalentReceivedSignal1}
 \begin{aligned}
 {\bf{y}}\left[ n \right] &= \left( {{{\bf{W}}^H}{\bf{\bar H\bar F}}} \right){\bf{s}}\left[ {n - {m_{\max }}} \right]+\\
 & \sum\limits_{i =  - {m_{{\rm{span}}}},i \ne 0}^{{m_{{\rm{span}}}}} {\left( {{{\bf{W}}^H}{\bf{\bar G}}\left[ i \right]{\bf{\bar F}}} \right){\bf{s}}\left[ {n - {m_{\max }} + i} \right]}  + {\bf{\tilde z}}\left[ n \right].
 \end{aligned}
 \end{equation}

 Since ${\bf s}\left[ n \right]$ is independent across $n$, the spectral efficiency of \eqref{DDAMEquivalentReceivedSignal1} in bps/Hz is
 \begin{equation}\label{delayDifferenceSE1}
 R = {\log _2}\left| {{{\bf{I}}_{{N_s}}} + {{\bf{W}}^H}{\bf{\bar H\bar F}}{{{\bf{\bar F}}}^H}{{{\bf{\bar H}}}^H}{\bf{W}}{{\left( {{{\bf{W}}^H}{\bf{CW}}} \right)}^{ - 1}}} \right|,
 \end{equation}
 where ${\bf{C}} =  {\sum\nolimits_{i =  - {m_{{\rm{span}}}},i \ne 0}^{{m_{{\rm{span}}}}} {{\bf{\bar G}}\left[ i \right]{\bf{\bar F}}} {{{\bf{\bar F}}}^H}{{{\bf{\bar G}}}^H}\left[ i \right] + {\sigma ^2}{\bf{I}}}$ is the interference-plus-noise covariance matrix. Our objective is to maximize the spectral efficiency by jointly optimizing the new transmit precoding matrix ${\bf{\bar F}}$ and the receive combining matrix $\bf{W}$. The problem can be formulated as
 \begin{equation}\label{originalOptimizationProblem}
 \begin{aligned}
 \left( {{\rm{P1}}} \right) \ \mathop {\max }\limits_{{\bf{\bar F}},{\bf{W}}} &\ \ {\log _2}\left| {{{\bf{I}}_{{N_s}}} + {{\bf{W}}^H}{\bf{\bar H\bar F}}{{{\bf{\bar F}}}^H}{{{\bf{\bar H}}}^H}{\bf{W}}{{\left( {{{\bf{W}}^H}{\bf{CW}}} \right)}^{ - 1}}} \right|\\
 {\rm{s.t.}}&\ \left\| {{\bf{\bar F}}} \right\|_F^2 \le P. \nonumber
 \end{aligned}
 \end{equation}

 Note that an efficient method for (P1) is the MSE-based method \cite{scaglione2002optimal,shi2011iteratively}. Specifically, for any given transmit precoding matrix ${\bf{\bar F}}$, the receive combining matrix is first optimized to minimize the MSE matrix ${\bf{E}}$, which is defined as the covariance matrix of the estimation error after receive combining matrix ${\bf{W}}$, i.e.,
 \begin{equation}\label{MSEMatrix}
 \begin{aligned}
 &{\bf{E}} = {{\mathbb E}}\left[ {\left( {{\bf{y}}\left[ n \right] - {\bf{s}}\left[ {n - {m_{\max }}} \right]} \right){{\left( {{\bf{y}}\left[ n \right] - {\bf{s}}\left[ {n - {m_{\max }}} \right]} \right)}^H}} \right]\\
 & = {{\bf{W}}^H}\left( {{\bf{\bar H\bar F}}{{{\bf{\bar F}}}^H}{{{\bf{\bar H}}}^H} + {\bf{C}}} \right){\bf{W}} - {{\bf{W}}^H}{\bf{\bar H\bar F}} - {{{\bf{\bar F}}}^H}{{{\bf{\bar H}}}^H}{\bf{W}} + {\bf{I}}.
  \end{aligned}
 \end{equation}
 It is observed that the MSE matrix is convex with respect to ${\bf {W}}$. By letting the first-order derivative of ${\bf{E}}$ with respect to ${\bf {W}}$ be equal to zero, we have
 \begin{equation}\label{optimalReceiveBeamforming}
 {{\bf{W}}^{\star}} = {\left( {{\bf{\bar H\bar F}}{{{\bf{\bar F}}}^H}{{{\bf{\bar H}}}^H} + {\bf{C}}} \right)^{ - 1}}{\bf{\bar H\bar F}},
 \end{equation}
 which is the well-known MMSE receiver or Wiener filter \cite{scaglione2002optimal,shi2011iteratively,palomar2003joint}. By substituting \eqref{optimalReceiveBeamforming} into \eqref{MSEMatrix}, the minimum MSE matrix can be given by
 \begin{equation}\label{MMSESolutionMSEMatrix}
 \begin{aligned}
 {\bf{E}_{\rm{MMSE}}} &= {\bf{I}} - {{{\bf{\bar F}}}^H}{{{\bf{\bar H}}}^H}{\left( {{\bf{\bar H\bar F}}{{{\bf{\bar F}}}^H}{{{\bf{\bar H}}}^H} + {\bf{C}}} \right)^{ - 1}}{\bf{\bar H\bar F}}\\
 &= {\left( {{\bf{I}} + {{{\bf{\bar F}}}^H}{{{\bf{\bar H}}}^H}{{\bf{C}}^{ - 1}}{\bf{\bar H\bar F}}} \right)^{ - 1}},
 \end{aligned}
 \end{equation}
 where the second equality holds due to the Woodbury matrix identity ${\left( {{\bf{A}} + {\bf{BCD}}} \right)^{ - 1}} = {{\bf{A}}^{ - 1}} - {{\bf{A}}^{ - 1}}{\bf{B}}{\left( {{{\bf{C}}^{ - 1}} + {\bf{D}}{{\bf{A}}^{ - 1}}{\bf{B}}} \right)^{ - 1}}{\bf{D}}{{\bf{A}}^{ - 1}}$.

 With the obtained receive combining matrix, the transmit precoding matrix is optimized to maximize the spectral efficiency. In the following, we first give the spectral efficiency expression under the MMSE receiver by showing that
 \begin{equation}\label{spectralEfficiencyUpperBound}
 \begin{aligned}
 &{\log _2}\left| {{{\bf{I}}_{{N_s}}} + {{\bf{W}}^H}{\bf{\bar H\bar F}}{{{\bf{\bar F}}}^H}{{{\bf{\bar H}}}^H}{\bf{W}}{{\left( {{{\bf{W}}^H}{\bf{CW}}} \right)}^{ - 1}}} \right|\\
 &= {\log _2}\left| {{{\bf{I}}_{{N_s}}} + {{{\bf{\bar F}}}^H}{{{\bf{\bar H}}}^H}{\bf{W}}{{\left( {{{\bf{W}}^H}{\bf{CW}}} \right)}^{ - 1}}{{\bf{W}}^H}{\bf{\bar H\bar F}}} \right|\\
 &= {\log _2}\left| {{{\bf{I}}_{{N_s}}} + {{{\bf{\bar F}}}^H}{{{\bf{\bar H}}}^H}{{\bf{C}}^{ - 1/2}}{{\bf{\Pi }}_{{{\bf{C}}^{1/2}}{\bf{W}}}}{{\bf{C}}^{ - 1/2}}{\bf{\bar H\bar F}}} \right|\\
 &\le {\log _2}\left| {{{\bf{I}}_{{N_s}}} + {{{\bf{\bar F}}}^H}{{{\bf{\bar H}}}^H}{{\bf{C}}^{ - 1}}{\bf{\bar H\bar F}}} \right|,
 \end{aligned}
 \end{equation}
 where the first equality is due to $\log \left| {{\bf{I}} + {\bf{AB}}} \right| $$= \log \left| {{\bf{I}} + {\bf{BA}}} \right|$, and ${{\bf{\Pi }}_{{{\bf{C}}^{1/2}}{\bf{W}}}} \triangleq {{\bf{C}}^{1/2}}{\bf{W}}{\left( {{{\bf{W}}^H}{\bf{CW}}} \right)^{ - 1}}{{\bf{W}}^H}{{\bf{C}}^{1/2}}$ is the orthogonal projector onto the space of ${{\bf{C}}^{1/2}}{\bf{W}}$. It is verified that the upper bound in \eqref{spectralEfficiencyUpperBound} is achieved when $\bf W$ is the MMSE receiver \cite{scaglione2002optimal}. As a result, the problem is equivalently formulated as
 \begin{equation}\label{transmitOptimizationProblem}
 \begin{aligned}
 \left( {{\rm{P2}}} \right) \ \mathop {\max }\limits_{{\bf{\bar F}}} &\ \ \ln \left| {{{\bf{I}}_{N_s}} + {{{\bf{\bar F}}}^H}{{{\bf{\bar H}}}^H}{{\bf{C}}^{ - 1}}{\bf{\bar H\bar F}}} \right|\\
 {\rm{s.t.}}&\ \ \left\| {{\bf{\bar F}}} \right\|_F^2 \le P. \nonumber
 \end{aligned}
 \end{equation}
 Let ${\bf{Q}} \in {{\mathbb {C}}^{{N_s} \times {N_s}}}$ be the auxiliary variable. Similar to \cite{hua2020intelligent}, (P2) can be transformed into the following problem
 \begin{equation}\label{transmitOptimizationEquivalentProblem}
 \begin{aligned}
 \left( {{\rm{P3}}} \right) \ \mathop {\max }\limits_{{\bf{Q}},{\bf{W}},{\bf{\bar F}}} &\ \ {\ln} \left| {\bf{Q}} \right| - {\rm{Tr}}\left( {{\bf{QE}}} \right) + {N_s}\\
 {\rm{s.t.}}&\ \ \left\| {{\bf{\bar F}}} \right\|_F^2 \le P,\\
 &\ \ {\bf{Q}} \succeq {\bf{0}}.\nonumber
 \end{aligned}
 \end{equation}
 where $\bf{E}$ in the objective function is given by \eqref{MSEMatrix}.

 Note that (P3) is still difficult to be solved due to the coupled variables ${\bf{Q}}$, ${\bf{W}}$, and ${\bf{\bar F}}$. To tackle this issue, the block coordinate descent (BCD) method is applied. Specifically, for given transmit precoding matrix ${\bf{\bar F}}$ and ${\bf{Q}}$, the optimal ${\bf{W}}$ can be obtained by letting the first-order derivative of the objective funtion with respect to $\bf{W}$ be equal to zero, given by \eqref{optimalReceiveBeamforming}, i.e., MMSE receiver. Similarly, for given transmit precoding matrix ${\bf{\bar F}}$ and the receive combining matrix ${\bf{W}}$, the optimal ${\bf{Q}}$ is obtained by letting the first-order derivative of the objective funtion with respect to ${\bf{Q}}$ be equal to zero, i.e.,
 \begin{equation}\label{optimalVariableQ}
 {{\bf{Q}}^{\star}} = {{\bf{E}}^{ - 1}}.
 \end{equation}

 Furthermore, for given ${\bf{W}}$ and $\bf{Q}$, by dropping the constant term and substituting \eqref{MSEMatrix} into the objective function of (P3), the transmit precoding matrix optimization problem reduces to
 \begin{equation}\label{transmitOptimizationEquivalentReducedProblem}
 \begin{aligned}
 \left( {{\rm{P4}}} \right) \ \mathop {\min }\limits_{{\bf{\bar F}}}&\  {\rm{Tr}}\left( {{\bf{Q}}{{\bf{W}}^H}{\bf{\bar H\bar F}}{{{\bf{\bar F}}}^H}{{{\bf{\bar H}}}^H}{\bf{W}}} \right)+ \\
 &\ \ {\rm{ Tr}}\left( {{\bf{Q}}{{\bf{W}}^H}\sum\limits_{i =  - {m_{{\rm{span}}}},i \ne 0}^{{m_{{\rm{span}}}}} {{\bf{\bar G}}\left[ i \right]{\bf{\bar F}}} {{{\bf{\bar F}}}^H}{{{\bf{\bar G}}}^H}\left[ i \right]{\bf{W}}} \right)-\\
 &\ \ {\rm{Tr}}\left( {{\bf{Q}}{{\bf{W}}^H}{\bf{\bar H\bar F}}} \right) - {\rm{Tr}}\left( {{\bf{Q}}{{{\bf{\bar F}}}^H}{{{\bf{\bar H}}}^H}{\bf{W}}} \right)\\
 {\rm{s.t.}}&\ \ \left\| {{\bf{\bar F}}} \right\|_F^2 \le P,\nonumber
 \end{aligned}
 \end{equation}
 which is a standard convex optimization problem, and can be solved by convex optimization tools, such as CVX. However, the structural property of the optimal solution can be obtained via the Lagrange dual method. Denote by $\beta \ge 0$ the Lagrange dual variable corresponding to the constraint. By following the similar procedure in \cite{shi2011iteratively}, the optimal solution to (P4) can be obtained as
 \begin{equation}\label{optimalTransmitPrecodingMatrix}
 \begin{aligned}
 &{{{\bf{\bar F}}}^{\star}} = \Big( {{{{\bf{\bar H}}}^H}{\bf{WQ}}{{\bf{W}}^H}{\bf{\bar H}} + } \Big.\\
 &{\Big. {\sum\limits_{i =  - {m_{{\rm{span}}}},i \ne 0}^{{m_{{\rm{span}}}}} {{{{\bf{\bar G}}}^H}\left[ i \right]{\bf{WQ}}{{\bf{W}}^H}{\bf{\bar G}}\left[ i \right]}  + \beta^{\star} {\bf{I}}} \Big)^{ - 1}}{{{\bf{\bar H}}}^H}{\bf{WQ}},
 \end{aligned}
 \end{equation}
 where the optimal dual variable $\beta^{\star}$ can be obtained via the bisection method as in \cite{shi2011iteratively}.

 Based on the obtained results, the overall MSE-based algorithm for solving (P1) is summarized in Algorithm~\ref{alg1}. Note that in each iteration, the variables are optimally obtained with other variables being fixed, and the resulting values of (P1) are non-decreasing. This thus guarantees the convergence of Algorithm~\ref{alg1}. Moreover, in Algorithm~\ref{alg1}, the complexity for step 3 is ${\cal O}\left( {M_r^3} \right)$ due to the matrix inversion, and that for step 4 is ${\cal O}\left( {N_s^3} \right)$. Step 5 has the complexity ${\cal O}\left( {{I_1}{L^3}M_t^3} \right)$, with $I_1$ denoting the number of iterations for bisection method. Thus, the overall complexity for Algorithm~\ref{alg1} is  ${\cal O}\left( {{I_2}\left( {M_r^3 + N_s^3 + {I_1}{L^3}M_t^3} \right)} \right)$, with $I_2$ denoting the number of iterations required by Algorithm~\ref{alg1} to converge.

 \begin{algorithm}[t]
 \caption{MSE-Based Algorithm for $(\rm P1)$}
 \label{alg1}
 \begin{algorithmic}[1]
 \STATE Initialize the precoding matrix satisfying $\left\| {{\bf{\bar F}}} \right\|_F^2 = P$.
 \REPEAT
 \STATE Obtain the receive combining matrix ${{\bf{W}}^{\star}}$ based on \eqref{optimalReceiveBeamforming}.
 \STATE Obtain the auxiliary variable ${{\bf{Q}}^{\star}}$ based on \eqref{optimalVariableQ}.
 \STATE Obtain the transmit precoding matrix ${{\bf{\bar F}}^{\star}}$ based on \eqref{optimalTransmitPrecodingMatrix}.
 \UNTIL the fractional increase of the objective function of (P1) is below a threshold.
 \end{algorithmic}
 \end{algorithm}

\section{Benchmark: MIMO-OFDM
considering ICI\\ and MIMO-OTFS}\label{sectionMIMOOFDMCommunication}

 \subsection{MIMO-OFDM Considering ICI}
 For MIMO-OFDM, let $K$ denote the number of sub-carriers, ${\bf{U}}\left[ k \right]$ and ${\bf{V}}\left[ k \right]$ denote the frequency-domain transmit precoding and receive combining matrices of sub-carrier $k$, respectively. Denote by $N_{\rm CP}$ the CP length of the OFDM scheme. Then the duration of each OFDM symbol including CP is $\left( {K + {N_{{\rm{CP}}}}} \right){T_s}$, and the number of OFDM symbols within each channel coherence time is ${N_{{\rm{OFDM}}}} = {N_c}/\left( {K + {N_{{\rm{CP}}}}} \right)$. Similarly, perfect CSI is needed for OFDM for transmit precoding and receive combining designs, which can be efficiently obtained via the compressed sensing scheme as in \cite{gaudio2022otfs}. With the frequency-domain transmit precoding matrix ${\bf{U}}\left[ k \right]$ and receive combining matrix ${\bf{V}}\left[ k \right]$, and by applying $K$-point discrete Fourier transform (DFT) to the received signal in \eqref{generalReceivedSignal}, the resulting frequency-domain signal is given by \eqref{ReceivedSignalOFDM}, shown at the top of this page, where ${{\bf{x}}_{\rm{f}}}\left[ q \right]$ denotes the information-bearing signal of sub-carrier $q$, with ${\mathbb E}[ {{{\bf{x}}_{\rm{f}}}[q]{\bf{x}}_{\rm{f}}^H[q]}] = {\bf{I}}$, ${\bf{z}}\left[ k \right] = \frac{1}{{\sqrt K }}\sum\nolimits_{n = 0}^{K - 1} {{\bf{z}}\left[ n \right]{e^{ - j\frac{{2\pi kn}}{K}}}}$ is the DFT of the AWGN, and ${{\bf{H}}_l}\left[ {q - k} \right] = \frac{1}{{ K }}\sum\nolimits_{n = 0}^{K - 1} {{{\bf{H}}_l}{e^{j2\pi {\nu _l}n{T_s}}}{e^{j\frac{{2\pi }}{K}\left( {q - k} \right)n}}}$ accounts for the ICI caused by multi-path $l$ \cite{jeon1999equalization}.
 \newcounter{mytempeqncnt2}
 \begin{figure*}
 \normalsize
 \setcounter{mytempeqncnt2}{\value{equation}}
 \begin{equation}\label{ReceivedSignalOFDM}
 \begin{aligned}
 {{\bf{y}}_{\rm{f}}}\left[ k \right] &= {{\bf{V}}^H}\left[ k \right]\frac{1}{{\sqrt K }}\sum\limits_{n = 0}^{K - 1} {{\bf{r}}\left[ n \right]{e^{ - j\frac{{2\pi kn}}{K}}}}  = \frac{1}{{\sqrt K }}{{\bf{V}}^H}\left[ k \right]\sum\limits_{n = 0}^{K - 1} {\left( {\sum\limits_{l = 1}^L {{{\bf{H}}_l}{e^{j2\pi {\nu _l}n{T_s}}}{\bf{x}}\left[ {n - {m_l}} \right]}  + {\bf{z}}\left[ n \right]} \right){e^{ - j\frac{{2\pi kn}}{K}}}} \\
 &= \frac{1}{{\sqrt K }}{{\bf{V}}^H}\left[ k \right]\sum\limits_{n = 0}^{K - 1} {\sum\limits_{l = 1}^L {{{\bf{H}}_l}{e^{j2\pi {\nu _l}n{T_s}}}\left( {\frac{1}{{\sqrt K }}\sum\limits_{q = 0}^{K - 1} {{\bf{U}}\left[ q \right]{{\bf{x}}_{\rm{f}}}\left[ q \right]{e^{j\frac{{2\pi }}{K}q\left( {n - {m_l}} \right)}}} } \right)} {e^{ - j\frac{{2\pi kn}}{K}}}}  + {{\bf{V}}^H}\left[ k \right]{\bf{z}}\left[ k \right]\\
 & = {{\bf{V}}^H}\left[ k \right]\sum\limits_{q = 0}^{K - 1} {\sum\limits_{l = 1}^L {{{\bf{H}}_l}\left[ {q - k} \right]{\bf{U}}\left[ q \right]{{\bf{x}}_{\rm{f}}}\left[ q \right]{e^{ - j\frac{{2\pi }}{K}q{m_l}}}} }  + {{\bf{V}}^H}\left[ k \right]{\bf{z}}\left[ k \right]\\
 & = {{\bf{V}}^H}\left[ k \right]\left( {\sum\limits_{l = 1}^L {{{\bf{H}}_l}\left[ 0 \right]{e^{ - j\frac{{2\pi }}{K}k{m_l}}}} } \right){\bf{U}}\left[ k \right]{{\bf{x}}_{\rm{f}}}\left[ k \right] + \underbrace {{{\bf{V}}^H}\left[ k \right]\sum\limits_{q \ne k}^{K - 1} {\sum\limits_{l = 1}^L {{{\bf{H}}_l}\left[ {q - k} \right]{e^{ - j\frac{{2\pi }}{K}q{m_l}}}} {\bf{U}}\left[ q \right]{{\bf{x}}_{\rm{f}}}\left[ q \right]} }_{{\rm{ICI}}} + {{\bf{V}}^H}\left[ k \right]{\bf{z}}\left[ k \right].
 \end{aligned}
 \end{equation}
 \hrulefill
 \end{figure*}

 In the following, the frequency-domain transmit precoding and receive combining matrices ${\bf{U}}\left[ k \right]$, ${\bf{V}}\left[ k \right]$ are designed to maximize the desired signal of sub-carrier $k$. Specifically, let the (reduced) SVD of $\sum\nolimits_{l = 1}^L {{{\bf{H}}_l}\left[ 0 \right]{e^{ - j\frac{{2\pi }}{K}k{m_l}}}}$ be ${\bf{R}}\left[ k \right]{\bf{\Sigma }}\left[ k \right]{{\bf{T}}^H}\left[ k \right]$, where ${\bf{ R}}\left[ k \right] \in {{\mathbb C}^{{M_r} \times {r_k}}}$, ${{{\bf{ T}}}\left[ k \right]} \in {{\mathbb C}^{{M_t} \times {r_k}}}$, and ${\bf{ \Sigma }}\left[ k \right] \in {{\mathbb C}^{{r_k} \times {r_k}}}$ contains $r_k$ positive singular values of $\sum\nolimits_{l = 1}^L {{{\bf{H}}_l}\left[ 0 \right]{e^{ - j\frac{{2\pi }}{K}k{m_l}}}}$. The receive combining and transmit precoding matrices are given by  ${{\bf{V}}\left[ k \right]} = {\bf{ R}}\left[ k \right]$ and ${\bf{U}}\left[ k \right] = \sqrt P {\bf{T}}\left[ k \right]/{\left\| {{\bf{T}}\left[ k \right]} \right\|_F}$. Then the resulting signal in \eqref{ReceivedSignalOFDM} reduces to
 \begin{equation}\label{reducedReceivedSignalOFDM}
 \begin{aligned}
 {{\bf{y}}_{\rm{f}}}\left[ k \right] =& \frac{{\sqrt P }}{ {{\left\| {{\bf{T}}\left[ k \right]} \right\|}_F}}{\bf{\Sigma }}\left[ k \right]{{\bf{x}}_{\rm{f}}}\left[ k \right] +\\
 &\sum\limits_{q \ne k}^{K - 1} {{\bf{\bar H}}\left[ {q - k} \right]{{\bf{x}}_{\rm{f}}}\left[ q \right]}  + {{\bf{R}}^H}\left[ k \right]{\bf{z}}\left[ k \right],
 \end{aligned}
 \end{equation}
 where ${\bf{\bar H}}\left[ {q - k} \right] = {{\bf{R}}^H}\left[ k \right]\sum\nolimits_{l = 1}^L {{{\bf{H}}_l}\left[ {q - k} \right]{e^{ - j\frac{{2\pi }}{K}q{m_l}}}} \frac{{\sqrt P {\bf{T}}\left[ q \right]}}{{{{\left\| {{\bf{T}}\left[ q \right]} \right\|}_F}}}$. Let ${\bf{\bar h}}_i^H\left[ {q - k} \right]$ and ${{\bf{r}}_i}\left[ k \right]$ denote the $i$th row and the $i$th column of ${\bf{\bar H}}\left[ {q - k} \right]$ and ${\bf{R}}\left[ k \right]$, respectively, $1 \le i \le {r_k}$. Thus, the resulting signal-to-interference-plus-noise ratio (SINR) of the $i$th data stream for sub-carrier $k$ is
 \begin{equation}\label{dataStreamSINR}
 \begin{aligned}
 {\gamma _{k,i}} &= \frac{{\frac{{P\Sigma _{i,i}^2\left[ k \right]}}{{\left\| {{\bf{T}}\left[ k \right]} \right\|_F^2}}}}{{{\mathbb E}\left[ {{{\left| {\sum\limits_{q \ne k}^{K - 1} {{\bf{\bar h}}_i^H\left[ {q - k} \right]{{\bf{x}}_{\rm f}}\left[ q \right]} } \right|}^2}} \right] + {\mathbb E}\left[ {{{\left| {{\bf{r}}_i^H\left[ k \right]{\bf{z}}\left[ k \right]} \right|}^2}} \right]}} \\
 &= \frac{{P\Sigma _{i,i}^2\left[ k \right]}}{{\left\| {{\bf{T}}\left[ k \right]} \right\|_F^2\left( {\sum\limits_{q \ne k}^{K - 1} {{{\left\| {{{{\bf{\bar h}}}_i}\left[ {q - k} \right]} \right\|}^2}}  + {\sigma ^2}} \right)}}.
 \end{aligned}
 \end{equation}

 By considering the CP overhead, the effective spectral efficiency in bps/Hz is given by
 \begin{equation}\label{spectralEfficiencyOFDM}
 {R_{{\rm{OFDM}}}} = \frac{{{N_c} - {N_{{\rm{OFDM}}}}{N_{{\rm{CP}}}}}}{{{N_c}}}\frac{1}{K}\sum\limits_{k = 0}^{K - 1} {\sum\limits_{i = 1}^{{r_k}} {{{\log }_2}\left( {1 + {\gamma _{k,i}}} \right)} }.
 \end{equation}
 \subsection{MIMO-OTFS}
 In this subsection, we study the MIMO-OTFS communication. Each OTFS frame contains $N$ symbols with the symbol duration $T$, and the bandwidth is divided into $M$ sub-carriers of spacing ${\Delta} f$, with $T\Delta f = 1$. The delay-Doppler domain representation of the doubly selective channel is
 \begin{equation}\label{DDDomainDoublySelectiveChannel}
 {\bf{H}}\left( {\tau ,\nu } \right) = \sum\limits_{l = 1}^L {{{\bf{H}}_l}\delta \left( {\tau  - {\tau _l}} \right)\delta \left( {\nu  - {\nu _l}} \right)},
 \end{equation}
 where ${\tau _l}$ denotes the delay of multi-path $l$. The delay and Doppler shift taps for multi-path $l$ are given by ${\tau _l} = {i_l}/\left(M\Delta f\right)$ and ${\nu _l} = {j_l}/\left(NT\right)$, respectively. For ease of comparison, we assume that $i_l$ and $j_l$ are integers. Let ${{\bf{X}}_{\rm{DD}}} \in {{\mathbb C}^{M \times N}}$ denote the information symbol matrix in the delay-Doppler plane. The time-domain transmitted signal ${\bf{s}} \in {{\mathbb C}^{MN \times 1}}$ can be expressed as ${\bf{s}} = \left( {{\bf{F}}_N^H \otimes {{\bf{G}}_{{\rm{tx}}}}} \right){{\bf{x}}_{{\rm{DD}}}}$  \cite{raviteja2018practical}, where with slight abuse of notations, ${{\bf{F}}_N}$ denotes the $N$-point DFT matrix, ${\bf{G}}_{\rm{tx}}$ denotes the transmit pulse-shaping matrix, and ${\bf{x}}_{\rm{DD}} = {\rm{vec}}\left( {{\bf{X}_{\rm{DD}}}} \right)$.

 Denote by ${\bf f} \in {\mathbb C}^{{M_t} \times 1}$ and ${\bf v} \in {\mathbb C}^{{M_r} \times 1}$ the transmit precoding and receive combining vectors, respectively. Let ${\bf{\Pi }} \in {{\mathbb C}^{MN \times MN}} ={\rm{circ}}\{ {{{\left[ {0,1,0, \cdots ,0} \right]}^T}} \}$ denote the permutation matrix, with ${\rm{circ}}\left\{ {\bf{x}} \right\}$ denoting the circular matrix whose first column is ${\bf x}$, and ${\bf{\Delta }} \in {{\mathbb C}^{MN \times MN}} = {\rm{diag\{ 1,}}{e^{j2\pi {\rm{/}}MN}}{\rm{,}} \cdots {\rm{,}}{e^{j2\pi (MN - 1){\rm{/}}MN}}{\rm{\} }}$. After removing the CP, the received signal is \cite{srivastava2022otfs}
 \begin{equation}\label{receivedSignalOTFS}
 {\bf{r}} = {\bf{Hs}} + {\bf{z}},
 \end{equation}
 where ${\bf{z}} \sim {\cal CN}\left({\bf{0}},{\sigma ^2}{{\bf{I}}_{MN}}\right)$ is the AWGN, and ${\bf{H}} = \sum\nolimits_{l = 1}^L {{{\tilde h}_l}{{\bf{\Pi }}^{{i_l}}}{{\bf{\Delta }}^{{j_l}}}}$, with ${\tilde h_l} = {{\bf{v}}^H}{{\bf{H}}_l}{\bf{f}}$. Moreover, the input-output relationship of the MIMO-OTFS system in the delay-Doppler domain is
 \begin{equation}\label{DDDomainReceivedSignalOTFS}
 {{\bf{y}}_{\rm{DD}}} = {{\bf{H}}_{\rm{DD}}}{{\bf{x}}_{\rm{DD}}} + {{\bf{z}}_{\rm{DD}}},
 \end{equation}
 where ${{\bf{H}}_{{\rm{DD}}}} = \left({{\bf{F}}_N} \otimes {{\bf{G}}_{{\rm{rx}}}}\right){\bf{H}}\left({\bf{F}}_N^H \otimes {{\bf{G}}_{{\rm{tx}}}}\right)$, with ${\bf{G}}_{\rm{rx}}$ denoting the receive pulse-shaping matrix, and ${{\bf{z}}_{\rm{DD}}} = \left({{\bf{F}}_N} \otimes {{\bf{G}}_{{\rm{rx}}}}\right){\bf{z}}$ is the delay-Doppler domain noise vector. In the following, the transmit precoding vector ${\bf f}$ and the receive combining vector ${\bf v}$ are optimized to maximize the channel gain. Let ${{\bf{\Psi }}_l} \triangleq {{\bf{\Pi }}^{{i_l}}}{{\bf{\Delta }}^{{j_l}}}$, the problem can be formulated as
 \begin{equation}\label{originalProblemOTFS}
 \begin{aligned}
 \left( {\rm{P{\text -}OTFS}} \right)\ \mathop {\max }\limits_{{\bf{f}},{\bf{v}}} &\ \ \left\| {\bf{H}} \right\|_F^2 = \Big\| {\sum\nolimits_{l = 1}^L {{{\tilde h}_l}{{\bf{\Psi }}_l}} } \Big\|_F^2\\
 {\rm{s.t.}}&\ \ \left\| {\bf{f}} \right\| = 1,\\
 &\ \ \left\|{\bf{v}} \right\| = 1.\nonumber
 \end{aligned}
 \end{equation}
 Problem (P-OTFS) is difficult to be directly solved since the optimization variables ${\bf f}$ and ${\bf v}$ are coupled with each other in the objective function. To tackle this problem, we propose an alternating optimization technique by iteratively optimizing the transmit precoding vector and the receive combining vector.
 \subsubsection{Transmit precoding vector optimization} For any given receive combining vector ${\bf v}$, the sub-problem of (P-OTFS) for optimizing the transmit precoding vector is written as
 \begin{equation}\label{subProblemOTFSTransmitPrecoding}
 \begin{aligned}
 \mathop {\max }\limits_{\bf{f}}&\ \ \Big\| {\sum\nolimits_{l = 1}^L {{{\tilde h}_l}{{\bf{\Psi }}_l}} } \Big\|_F^2\\
 {\rm{s.t.}}&\ \ \left\| {\bf{f}} \right\| = 1.
 \end{aligned}
 \end{equation}
 The objective function can be expressed as
 \begin{equation}\label{objectiveFunctionTransmitPrecoding}
 \begin{aligned}
 &\Big\| {\sum\nolimits_{l = 1}^L {{{\tilde h}_l}{{\bf{\Psi }}_l}} } \Big\|_F^2 = {\rm{Tr}}\left( {\Big( {\sum\nolimits_{l = 1}^L {\tilde h_l^*{\bf{\Psi }}_l^H} } \Big)\Big( {\sum\nolimits_{l' = 1}^L {{{\tilde h}_{l'}}{{\bf{\Psi }}_{l'}}} } \Big)} \right)\\
 &= {{\bf{f}}^H}\left( {\sum\nolimits_{l = 1}^L {\sum\nolimits_{l' = 1}^L {{\bf{H}}_l^H{\bf{v}}{{\bf{v}}^H}{{\bf{H}}_{l'}}{\rm{Tr}}\left( {{\bf{\Psi }}_l^H{{\bf{\Psi }}_{l'}}} \right)} } } \right){\bf{f}}.
 \end{aligned}
 \end{equation}
 Let ${\bf{\Lambda }} \triangleq \sum\nolimits_{l = 1}^L {\sum\nolimits_{l' = 1}^L {{\bf{H}}_l^H{\bf{v}}{{\bf{v}}^H}{{\bf{H}}_{l'}}{\rm{Tr}}\left( {{\bf{\Psi }}_l^H{{\bf{\Psi }}_{l'}}} \right)} }$. The optimal transmit precoding vector to maximize \eqref{objectiveFunctionTransmitPrecoding} can be obtained by taking the eigenvector corresponding to the maximum eigenvalue of the matrix ${\bf{\Lambda }} $.
 \subsubsection{Receive combining vector optimization}
 For any given transmit precoding vector, the sub-problem of (P-OTFS) for optimizing the receive combining vector is
 \begin{equation}\label{subProblemOTFSReceiveCombining}
 \begin{aligned}
 \mathop {\max }\limits_{\bf{v}}&\ \ \Big\| {\sum\nolimits_{l = 1}^L {{{\tilde h}_l}{{\bf{\Psi }}_l}} } \Big\|_F^2\\
 {\rm{s.t.}}&\ \ \left\| {\bf{v}} \right\| = 1.
 \end{aligned}
 \end{equation}
 Similarly, the objective function can be expressed as
 \begin{equation}\label{objectiveFunctionReceiveCombining}
 \begin{aligned}
 \Big\| {\sum\nolimits_{l = 1}^L {{{\tilde h}_l}{{\bf{\Psi }}_l}} } \Big\|_F^2 &= {\rm{Tr}}\left( {\Big( {\sum\nolimits_{l = 1}^L {{{\tilde h}_l}{{\bf{\Psi }}_l}} } \Big)\Big( {\sum\nolimits_{l' = 1}^L {\tilde h_{l'}^*{\bf{\Psi }}_{l'}^H} } \Big)} \right)\\
 &= {{\bf{v}}^H}{\bf{\Gamma v}},
 \end{aligned}
 \end{equation}
 where ${\bf{\Gamma }} \triangleq \sum\nolimits_{l = 1}^L {\sum\nolimits_{l' = 1}^L {{{\bf{H}}_l}{\bf{f}}{{\bf{f}}^H}{\bf{H}}_{l'}^H{\rm{Tr}}\left( {{{\bf{\Psi }}_l}{\bf{\Psi }}_{l'}^H} \right)} } $. Then the optimal receive combining vector can be obtained by taking the eigenvector corresponding to the maximum eigenvalue of the matrix ${\bf{\Gamma }} $.

 As such, problem (P-OTFS) can be solved by iteratively optimizing the transmit precoding and receive combining matrices. Note that the resulting value of (P-OTFS) is non-decreasing over each iteration, thus guaranteeing the convergence. With the obtained transmit precoding and receive combining matrices, and by considering the rectangular transmit and receive pulse shaping waveforms, the effective spectral efficiency of MIMO-OTFS in bps/Hz is given by  \cite{pandey2021low}
 \begin{equation}\label{spectralEfficiencyOTFS}
 {R_{{\rm{OTFS}}}} = \frac{1}{{MN + {N'_{{\rm{CP}}}}}}{\log _2}\left| {{\bf{I}} + \bar P{{\bf{H}}_{{\rm{DD}}}}{\bf{H}}_{{\rm{DD}}}^H} \right|,
 \end{equation}
 where ${N'_{{\rm{CP}}}}$ denotes the CP length of the OTFS scheme.

 Last, the guard interval overheads of DDAM, OFDM and OTFS are compared. By exploiting the path-invariant property, one new finding is that to circumvent the ISI across different path invariant blocks, DDAM only needs a guard interval of length $2{m_{\max }}$ for each path invariant block $\bar T$, by noting that the resulting signal is at most delayed by ${m_{\max }} + {m_{{\rm{span}}}} \approx 2{m_{\max }}$, as can be seen in \eqref{DDAMReceivedSignal}. Compared to that revealed in \cite{lu2022delay}, where a guard interval of length $2{m_{\max }}$ is required for each channel coherence block $T_c$, this further reduces the guard interval overhead, since ${\bar T} \gg {T_c}$. Thus, the guard interval overhead for DDAM is $2{m_{\max }}/{\bar N}$. Besides, by choosing the CP of length ${m_{\max }}$, the CP overhead of OFDM is given by ${N_{{\rm{OFDM}}}}{N_{{\rm{CP}}}}/{N_c} = {m_{\max }}/\left( {{m_{\max }} + K} \right)$. It is observed that compared to OFDM, the proposed DDAM transmission is able to substantially reduce the CP overhead since one path invariant block in general contains multiple OFDM symbols, i.e., ${\bar N} \gg {m_{\max }} + K$. Moreover, by choosing ${N'_{{\rm{span}}}} = {m_{\max }}$, the CP overhead for OTFS is ${m_{\max }}/\left( {MN + {m_{\max }}} \right)$.

\section{Numerical Results}\label{sectionNumericalResults}
 In this section, numerical results are provided to validate the performance of the proposed DDAM transmission. We consider a mmWave system with the carrier frequency and the total bandwidth given by $f = 28$ GHz and $B = 100$ MHz, respectively. The noise power spectrum density is $N_0 = -174$ dBm/Hz. Both the BS and the UE are equipped with the uniform linear array (ULA), where the adjacent antenna elements are separated by half-wavelength. The number of significant multi-paths is $L =3$ \cite{song2019efficient}. The delays are uniformly distributed in $\left[ {0,{\tau_{\rm{max}}}} \right]$, with $\tau_{\rm{max}} = 400$ ns. Thus, the upper bound of the channel delay spread is ${{\tilde m}_{\max }} = {m_{\max }} = {\tau _{\max }}B = 40$. The channel matrix of multi-path $l$ is ${{\bf{H}}_l} = {\alpha _l}{{\bf{a}}_R}\left( {{\phi _l}} \right){\bf{a}}_T^H\left( {{\varphi _l}} \right)$, and the AoAs $\left\{ {\phi _l} \right\}_{l = 1}^L$ and AoDs $\left\{ {\varphi _l} \right\}_{l = 1}^L$ are equally spaced in $[ - {60^ \circ },{60^ \circ }]$. The complex-valued channel coefficients $\left\{ {{\alpha _l}} \right\}_{l = 1}^L$ are generated based on the model given in \cite{akdeniz2014millimeter}, which is developed by using the channel measurements for mmWave communications. The generation of Doppler frequencies follows the Jakes' formula ${\nu _l} = {\nu _{\max }}\cos \left( {{\theta _l}} \right)$, where $\nu _{\max }$ is the maximum Doppler frequency determined by UE's velocity and ${\theta _l}$ is uniformly distributed in $\left[ { - \pi ,\pi } \right]$ \cite{raviteja2018practical}. Unless otherwise stated, the number of receive antennas and data streams are $M_r = N_s = 2$, the total available power is $P =30$ dBm, and the velocity is $v = 180$ km/h. By setting $\zeta = 0.1$, the channel coherence time is ${T_c} = 0.1/{\nu_{\max}}= 0.1c/vf \approx 0.0214$ ms, within which the total number of signal samples is ${N_c} = B{T_c} = 2142$. Moreover, the path invariant time can be obtained by ensuring $v{\bar T} \le c/B$. For the benchmarking scheme of OFDM, the number of sub-carriers is $K = 512$, and a CP of length ${N_{{\rm{CP}}}} = {{\tilde m}_{\max }} = 40$ is used. Beside, for the benchmarking scheme of OTFS, the number of sub-carriers and time-slots are $M = 512$ and $N=8$, respectively.

 \begin{figure}[!t]
 \centering
 \centerline{\includegraphics[width=3.5in,height=2.625in]{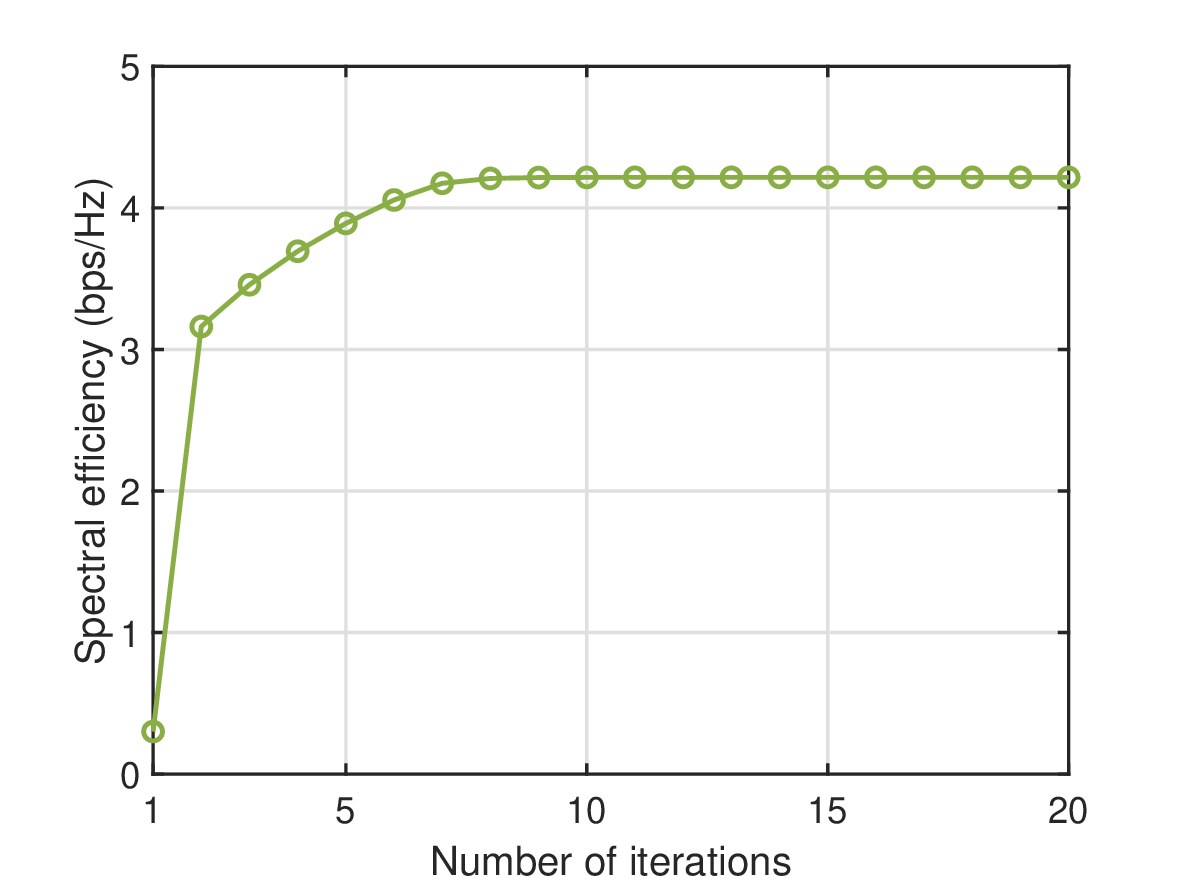}}
 \caption{Convergence behaviour of Algorithm~\ref{alg1}.}
 \label{BCDConvergencePerformance}
 \end{figure}

 Fig.~\ref{BCDConvergencePerformance} gives the convergence behaviour of Algorithm~\ref{alg1}. The number of transmit antennas is $M_t=64$. It is observed that the MSE-based algorithm monotonically converges within few iterations, which validates the analysis in Section \ref{sectionJointPrecodingCombining}.

 \begin{figure}[!t]
 \centering
 \centerline{\includegraphics[width=3.5in,height=2.625in]{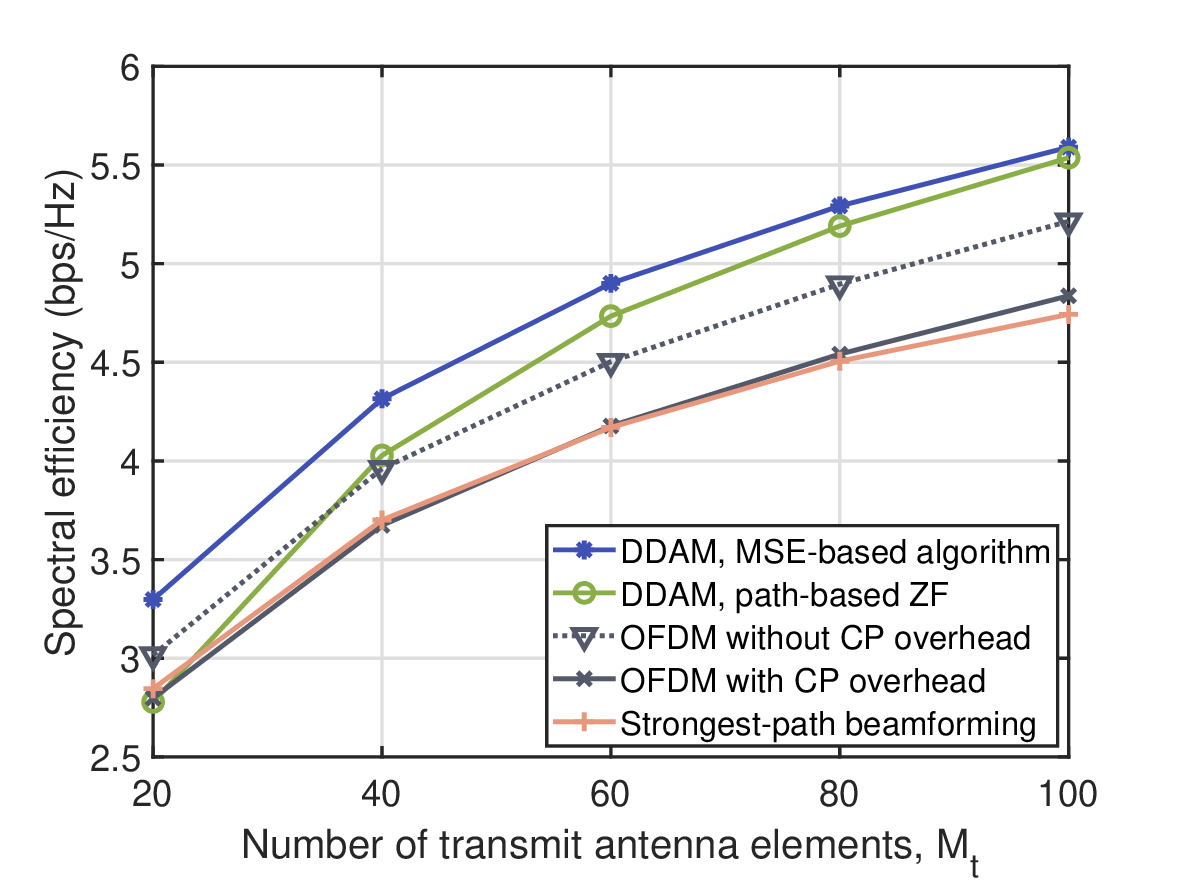}}
 \caption{Spectral efficiency versus the number of transmit antennas.}
 \label{spectralEfficiencyVersusTransmitAntennaNumber}
 \end{figure}

 Fig.~\ref{spectralEfficiencyVersusTransmitAntennaNumber} shows the spectral efficiency versus the number of transmit antennas $M_t$. For comparison, the single-carrier benchmarking scheme of strongest-path beamforming is considered, where the transmit precoding and receive combining matrices are designed only based on the single dominant path, and the single delay and Doppler frequency along the dominant path can be easily addressed by standard synchronization techniques at the receiver. In this scheme, the transmitted signal is ${\bf{x}}\left[ n \right] = {\bf{Fs}}\left[ n \right]$. The optimization of transmit precoding and receive combining matrices is similar to (P1). It is observed that as the number of antennas increases, the MSE-based and path-based ZF DDAM transmissions give comparable performance, and they outperform the conventional OFDM. This is mainly attributed to the following two reasons. Firstly, compared to OFDM, the proposed DDAM transmission significantly saves the guard interval overhead. To be specific, the guard interval overhead of DDAM is $2{{\tilde m}_{\max }}/{\bar N} = 0.0013 \%$, whereas the CP overhead of OFDM is ${{\tilde m}_{\max }}/\left( {{{\tilde m}_{\max }} + K} \right) = 7.25 \%$. Secondly, OFDM is susceptible to the detrimental ICI caused by the UE mobility, while the single-carrier DDAM shows tolerance to the Doppler effect by utilizing path-based Doppler-compensation. This is reflected by the observation that even when the impact of CP overhead is removed, the proposed DDAM transmission is still superior to OFDM. It is also observed that the two DDAM transmission schemes outperform the strongest-path beamforming scheme, which is expected since DDAM makes full use of all the multi-path signal components, as can be seen in \eqref{DDAMReceivedSignal}, whereas the strongest-path beamforming scheme only uses the dominant path component as the desired signal.

 \begin{figure}[!t]
 \centering
 \centerline{\includegraphics[width=3.5in,height=2.625in]{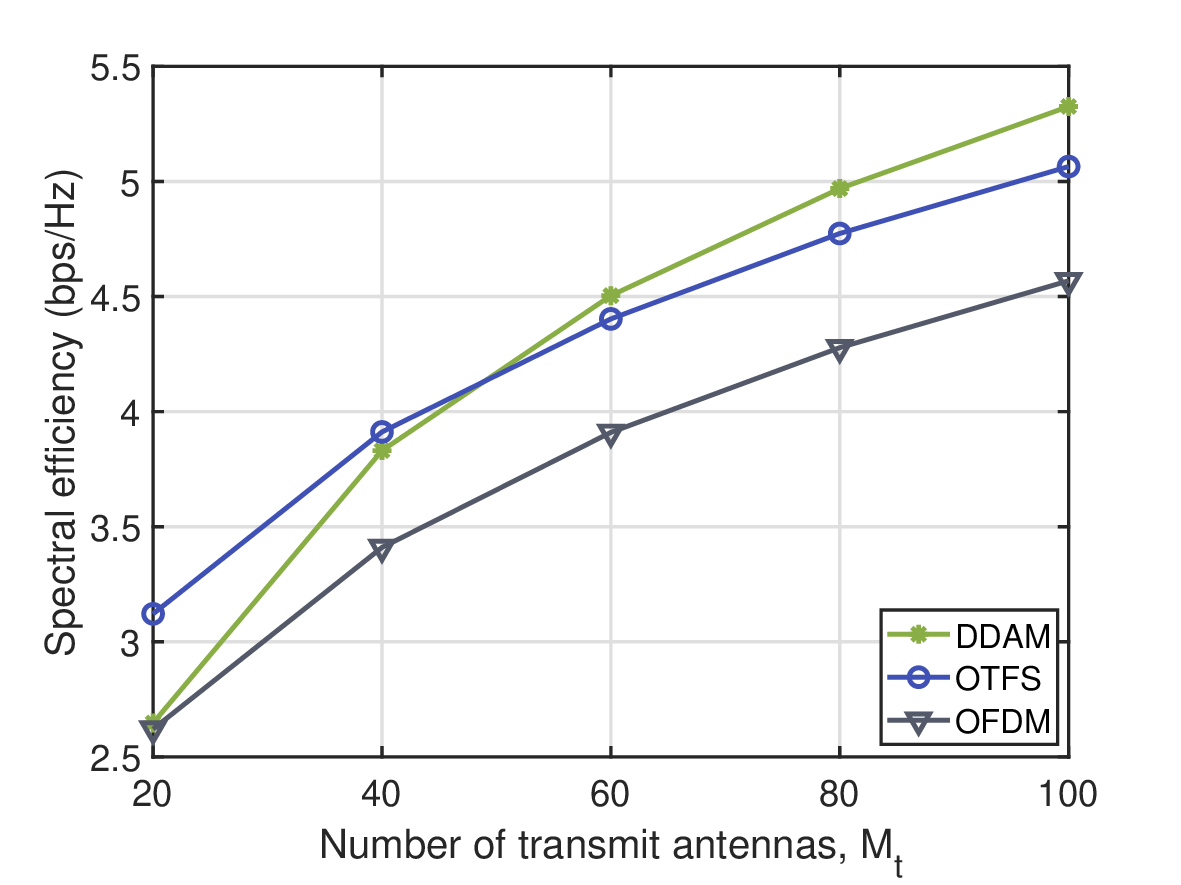}}
 \caption{Comparison of spectral efficiency for DDAM, OTFS, and OFDM.}
 \label{spectralEfficiencyDDAMOFDMOTFS}
 \end{figure}

 Fig.~\ref{spectralEfficiencyDDAMOFDMOTFS} shows a further comparison of spectral efficiency for DDAM, OTFS, and OFDM. The velocity of the UE is $v = 500$ km/h. It is observed that as the number of transmit antennas increases, the proposed DDAM technique outperforms OTFS, and the two schemes significantly outperform the conventional OFDM. This is expected since both DDAM and OTFS are robust to the Doppler effect caused by the high mobility. Moreover, compared to OTFS, DDAM has the additional advantages of reduced receiver complexity and processing, rendering it quite appealing for systems with multi-path sparsity and high spatial dimensions.

 \begin{figure}[!t]
 \centering
 \centerline{\includegraphics[width=3.5in,height=2.625in]{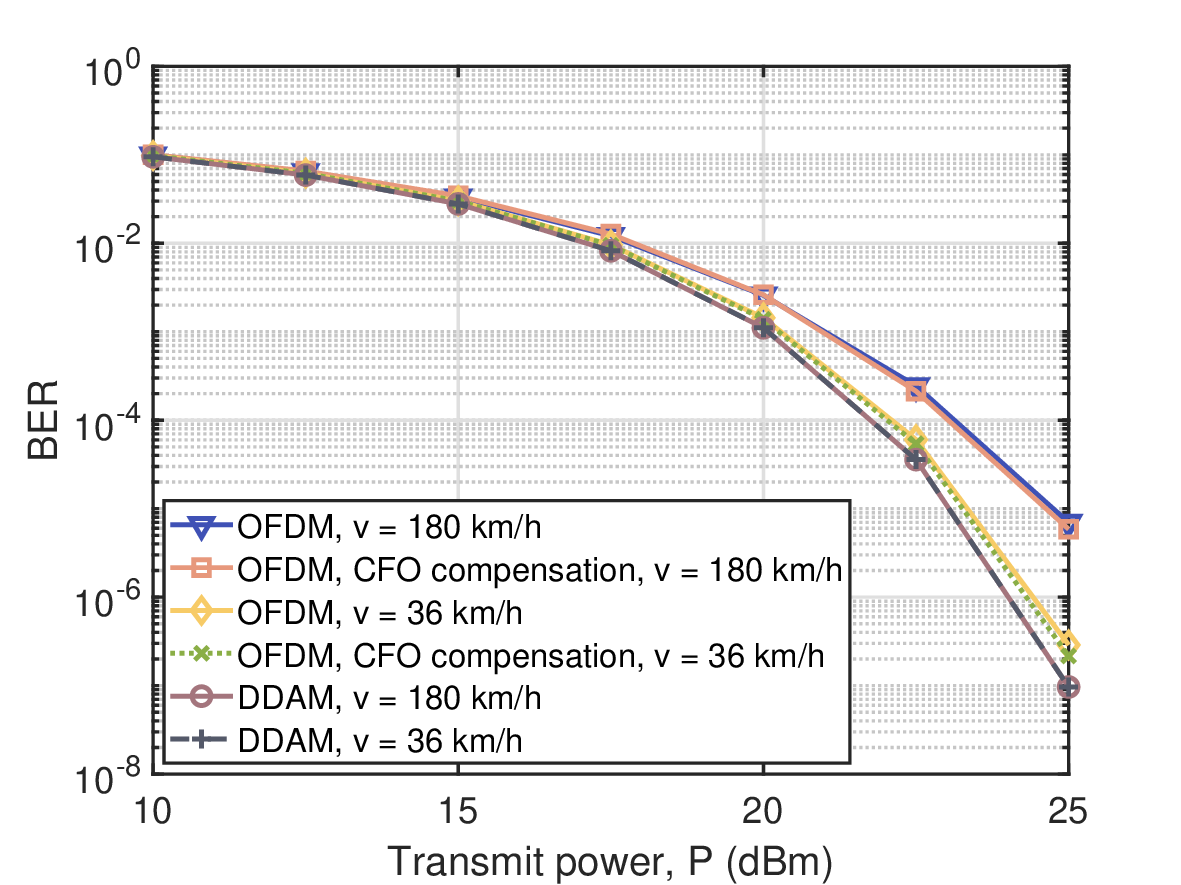}}
 \caption{BER versus the transmit power for the proposed DDAM, OFDM, and OFDM adopting CFO compensation.}
 \label{BERComparisionDDAMVersusOFDM}
 \end{figure}

 Fig.~\ref{BERComparisionDDAMVersusOFDM} shows the BER performance comparison of the proposed path-based ZF DDAM transmission, the conventional OFDM, and OFDM adopting CFO compensation, by adopting the 128 quadrature amplitude modulation (QAM). The number of transmit antennas and data streams are set as $M_t = 256$ and $N_s = 1$, respectively. For OFDM adopting CFO compensation, the common Doppler frequency compensation is ${e^{ - j2\pi {\nu _1}n{T_s}}}$. In particular, the ICI terms in \eqref{ReceivedSignalOFDM} can be approximately modeled as Gaussian noise for large number of sub-carriers $K$, due to the central limit theorem \cite{cheon2002effect}. Denote by ${P_e}\left( \gamma  \right)$ the BER versus $\gamma$ over the AWGN channel for QAM modulation, thus the BER expression of OFDM can be approximately given by \cite{xia2001precoded}
 \begin{equation}\label{OFDMBERExpression}
 \begin{aligned}
 {P_{e,{\rm{OFDM}}}} = \frac{1}{K}\sum\nolimits_{k = 1}^K {{P_e}\left( {\frac{{K{\gamma _k}}}{{K + {{\tilde m}_{\max }}}}} \right)},
 \end{aligned}
 \end{equation}
 where ${\gamma _k}$ denotes the effective SNR and can be obtained from \eqref{dataStreamSINR} by letting $i=1$. On the other hand, the path-based ZF DDAM transmission achieves the transformation of the doubly selective channel into time-invariant ISI-free channel, as shown in Section \ref{sectionPathBasedZFPrecoding}, whose BER expression is given by ${P_e}\left( \gamma _{\rm{ZF}} \right)$, with $\gamma _{\rm{ZF}} = P\tilde \Sigma _{1,1}^2/{\sigma ^2}$. It is observed that while CFO compensation is effective for OFDM when all multi-paths have the identical Doppler frequency shift \cite{cho2010mimo}, for the considered general scenario where multi-paths have different Doppler shifts, CFO compensation does not work well. This is expected since the common Doppler frequency compensation cannot compensate for the distinct Doppler shifts of all the different multi-paths, even when perfect CSI is available for OFDM. It is also observed that the DDAM transmission yields a better BER performance than the conventional OFDM, even for the small UE velocity $v = 36$ km/h, which corresponds to the case of a smaller difference between Doppler shifts of different paths. Besides, as the UE velocity increases from 36 km/h to 180 km/h, the performance gap becomes more pronounced. This is because, on one hand, OFDM requires larger CP overhead than DDAM, which is manifested by the coefficient $K/\left( {K + {{\tilde m}_{\max }}} \right)$ in \eqref{OFDMBERExpression}. On the other hand, the increase of velocity renders the Doppler effect and the ICI more severe for OFDM, thus deteriorating its BER performance. By contrast, DDAM exhibits substantial resilience to the Doppler effect by completely eliminating it, which is quite attractive for the high-speed scenarios.

 \begin{figure}[!t]
 \centering
 \centerline{\includegraphics[width=3.5in,height=2.625in]{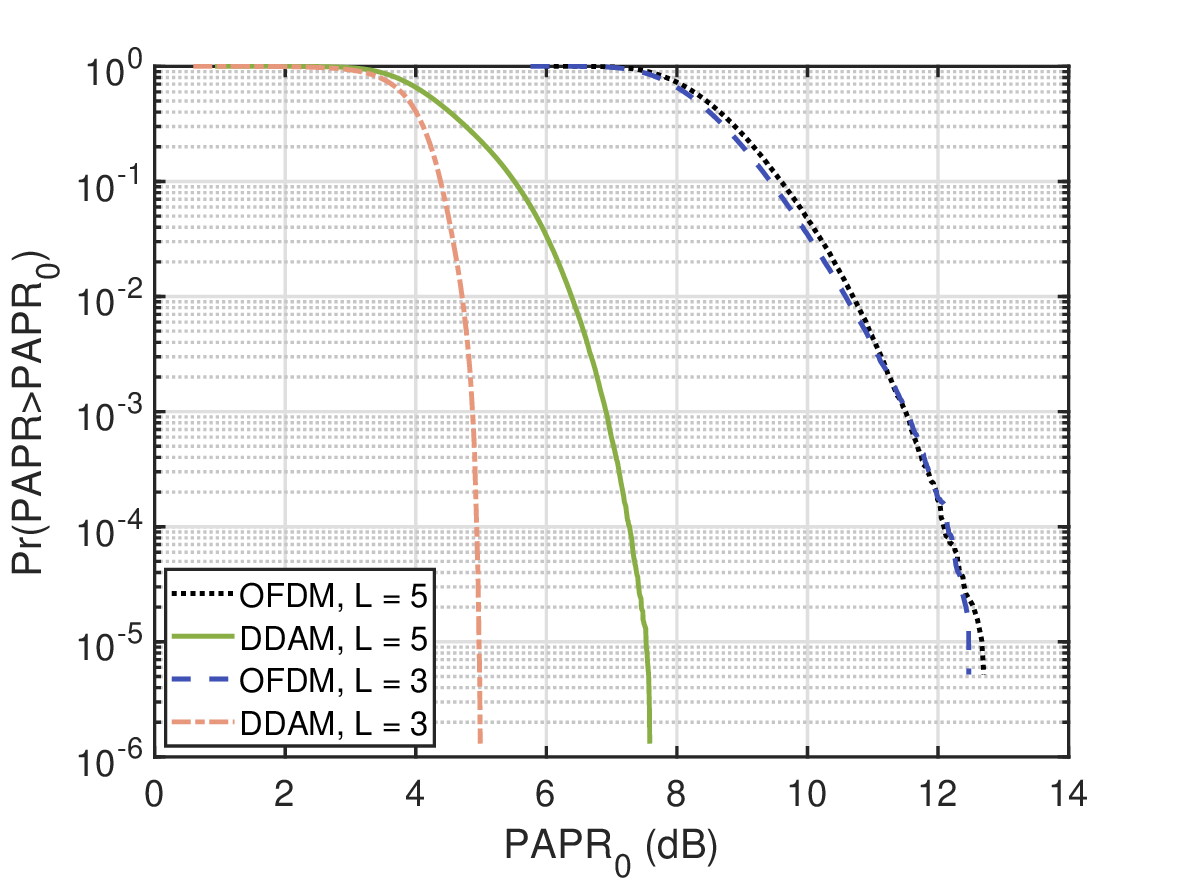}}
 \caption{Complementary cumulative distribution function of the PAPR for the proposed DDAM and the conventional OFDM.}
 \label{PAPRComparison}
 \end{figure}

 To evaluate the PAPR performance of the transmitted signals of the path-based ZF DDAM transmission and OFDM, the metric of the complementary cumulative distribution function (CCDF) is used in Fig.~\ref{PAPRComparison}. The number of transmit antennas is $M_t =128$, and the modulation scheme is 128 QAM. It is observed from Fig.~\ref{PAPRComparison} that the proposed DDAM transmission achieves significant PAPR reduction than the conventional OFDM for both $L=3$ and $L=5$. This is expected since the transmitted signal of OFDM is the superposition of $K$ sub-carriers. By contrast, the transmitted signal of the DDAM transmission is on a per-path basis, as can be seen in \eqref{DDAMTransmitSignal}, which only depends on the number of multi-paths $L \ll K$, and thus each transmit antenna experiences a lower PAPR. The above results show the superiority of DDAM to OFDM in terms of spectral efficiency, BER, and PAPR.

\begin{figure}[!t]
 \centering
 \centerline{\includegraphics[width=3.5in,height=2.625in]{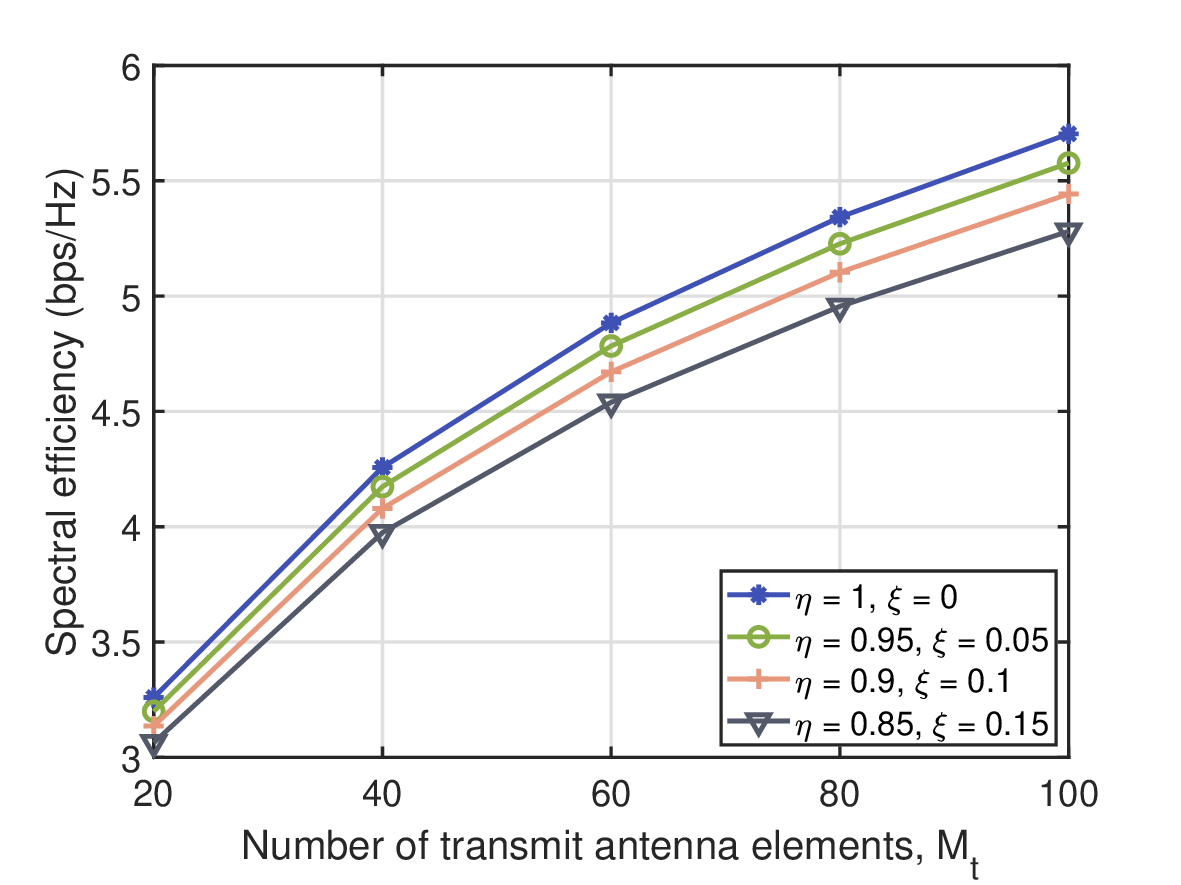}}
 \caption{Spectral efficiency versus the number of transmit antennas under imperfect knowledge of delay and Doppler shifts.}
 \label{imperfectKnowledgeofDelayDoppler}
 \end{figure}

 Last, Fig.~\ref{imperfectKnowledgeofDelayDoppler} shows the spectral efficiency versus the number of transmit antennas under the imperfect knowledge of delay and Doppler shifts. Let ${\hat m_l}$ and ${\hat \nu_l}$ denote the estimated delay and Doppler shift of multi-path $l$, respectively, and ${\bf{e}} \triangleq \left[{e_1}, \cdots ,{e_L}\right]^T$, where ${e_l}$ is the indicator variable for accurately estimating the delay of multi-path $l$, i.e., if ${\hat m_l} = {m_l}$, ${e_l} = 1$; otherwise, ${e_l} = 0$. The delay estimation accuracy is defined as $\eta  = {\left\| {\bf{e}} \right\|^2}/L$. Note that with perfect delay estimation, we have ${\left\| {\bf{e}} \right\|^2} = L$ and $\eta  = 1$, and a smaller $\eta$ corresponds to a larger delay estimation error. Besides, the imperfect Doppler shift is modelled as ${\hat \nu _l} = {\nu _l} + {\tilde \nu _l}$, where ${\tilde \nu _l} \sim {\cal CN}(0,{\xi ^2}\nu _{\max }^2)$ denotes the estimation error of Doppler shifts. It is observed that the spectral efficiency of DDAM slightly decreases as the delay and Doppler estimation error increase, which implies that DDAM has certain robustness to the imperfect knowledge of delay and Doppler shifts.

\section{Conclusion}\label{sectionConclusion}
 This paper proposed a novel Doppler-ISI double mitigation technique termed DDAM for time-variant frequency-selective massive MIMO communications. We first showed that DDAM can transform the time-variant frequency-selective channels into time-invariant ISI-free channels by applying path-based ZF precoding and receive combining, followed by the derivation of sufficient and/or necessary conditions to achieve such a transformation. In particular, when the number of BS antennas is much larger than that of channel paths, the time-invariant ISI-free channel was enabled with the simple delay-Doppler compensation and path-based MRT beamforming. Furthermore, by tolerating some residual ISI, the MSE-based method was developed for solving the spectral efficiency maximization problem. Numerical results were provided to compare the proposed DDAM technique with MIMO-OTFS, MIMO-OFDM without or with CFO compensation, and beam alignment along the dominant path.

 Though DDAM has promising advantages, it faces several practical challenges. For example, DDAM is most effective for mmWave/THz with sparse and separable multi-paths, while is difficult to be implemented for sub-6 GHz bands. Besides, the low-complexity hybrid beamforming design, as well as the diversity order and detection methods for DDAM require in-depth study in the future. Moreover, to avoid the channel aging arising from channel feedback \cite{truong2013effects,papazafeiropoulos2016impact}, the reserve channel estimation can be performed at the transmitter, by exploiting the path reciprocity of the physical channels between downlink and uplink, including the angle, delay, and Doppler shift, as revealed in \cite{wang2019overview}. CSI acquisition of the individual path and the impact of CSI error on DDAM performance are important to investigate in the future.

\begin{appendices}
\section{Proof of Proposition~\ref{ZFInfeasibilityProposition}}\label{ProofZFInfeasibilityProposition}
 Proposition~\ref{ZFInfeasibilityProposition} can be shown by counting the number of equations, $N_e$, and that of free design variables, $N_v$, in \eqref{ZFDDAMCondition}. When ${N_v} \ge {N_e}$, the system is said to be proper, otherwise, it is called an improper system \cite{yetis2010feasibility,razaviyayn2012degrees}. It is verified that improper systems are infeasible, though proper systems do not guarantee feasibility \cite{razaviyayn2012degrees,wang2014subspace}. In the following, we first determine the number of bilinear equations corresponding to the ZF conditions in \eqref{ZFDDAMCondition}. Denote by ${{\bf{w}}_j} \in {{\mathbb{C}}^{{M_r} \times 1 }}$ and ${{\bf{f}}_{l,k}}\in {{\mathbb{C}}^{{M_t} \times 1 }}$ the $j$th column of the receive combining matrix ${\bf{W}}$ and the $k$th column of the transmit precoding matrix ${{\bf{F}}_l}$, respectively, where $j,k= 1, \cdots,{N_s}$. Then the ZF conditions \eqref{ZFDDAMCondition:sub1} can be equivalently expressed as
 \begin{equation}\label{vectorZFDDAMCondition}
{\bf{w}}_j^H{{\bf{H}}_{l}}{{\bf{f}}_{l',k}} = 0,\ \forall l \ne l'\ {\rm and}\  \forall j,k.
 \end{equation}
 The number of equations in \eqref{vectorZFDDAMCondition} is
 \begin{equation}\label{numberofEquation}
 {N_e} = \sum\limits_{l = 1}^L {\sum\limits_{l' \ne l}^L {{N_s} \times {N_s}} }  = L\left( {L - 1} \right)N_s^2.
 \end{equation}
 Next, we count the number of free design variables in ${{\bf{F}}_l}$ and ${\bf{W}}$. By following the similar procedure in \cite{yetis2010feasibility}, the total number of variables to be freely designed is given by
 \begin{equation}\label{numberofVariables}
 \begin{aligned}
 {N_v} &= {N_s}\left( {{M_r} - {N_s}} \right) + L{N_s}\left( {{M_t} - {N_s}} \right) \\
 &= {N_s}\left( {L{M_t} + {M_r}} \right) - \left( {L + 1} \right)N_s^2.
 \end{aligned}
 \end{equation}

 Recall that the system is improper when ${N_v} < {N_e}$, i.e., $L{M_t} + {M_r} < \left( {{L^2} + 1} \right){N_s}$, and improper systems are infeasible \cite{razaviyayn2012degrees,wang2014subspace}, which implies that \eqref{ZFNecessaryCondition} is a necessary condition for the ZF conditions in \eqref{ZFDDAMCondition} to be feasible. The proof of Proposition~\ref{ZFInfeasibilityProposition} is thus completed.

\section{Proof of Proposition~\ref{ZFSufficientConditionProposition}}\label{ProofZFSufficientConditionProposition}
 To show Proposition~\ref{ZFSufficientConditionProposition}, the path-based precoding matrices $\{ {{\bf{F}}_l}\} _{l = 1}^L$ are designed so that
 \begin{subequations}\label{pathBasedZFBeamforming}
 \begin{align}
 &{{\bf{H}}_l}{{\bf{F}}_{l'}} = {{\bf{0}}_{{M_r} \times {N_s}}}, \ \forall l \ne l', \label{pathBasedZFBeamforming:sub1}\\
 &{\rm{rank}}\left( {{{\bf{F}}_l}} \right) = {N_s},\ \forall l,\label{pathBasedZFBeamforming:sub2}
 \end{align}
 \end{subequations}
 which is a sufficient but not necessary condition of \eqref{ZFDDAMCondition}. The above ZF conditions in \eqref{pathBasedZFBeamforming:sub1} can be compactly written as ${\bf{\tilde H}}_{l'}^H{{\bf{F}}_{l'}} = {{\bf{0}}_{\left( {L - 1} \right){M_r} \times {N_s}}}$, where ${{{\bf{\tilde H}}}_{l'}} \in {{\mathbb C}^{{M_t} \times \left( {L - 1} \right){M_r}}} \triangleq \left[ {{\bf{H}}_1^H, \cdots ,{\bf{H}}_{l' - 1}^H,{\bf{H}}_{l' + 1}^H, \cdots ,{\bf{H}}_L^H} \right]$. As a result, the ZF conditions are feasible as long as ${{{\bf{\tilde H}}}_{l'}}$ has a $N_s$-dimensional null-space, which is true almost surely when ${M_t} \ge \left( {L - 1} \right){M_r} + {N_s}$. This thus completes the proof of Proposition~\ref{ZFSufficientConditionProposition}.

 \section{Proof of Proposition~\ref{ReceivedSNRMRT2OptimalUpperBound}}\label{ProofReceivedSNRMRT2OptimalUpperBound}
 To maximize the SNR in \eqref{ReceivedSNRMRT2UpperBound}, the power allocation $\left\{ {{p_l}} \right\}_{l = 1}^L$ is optimized, and the optimization problem can be equivalently formulated as
 \begin{equation}
 \begin{aligned}
 \left( {\rm{P{\text -}MRT}} \right)\ \mathop {\max }\limits_{\left\{ {{p_l}} \right\}_{l = 1}^L}&\ \ \sum\limits_{l = 1}^L {\sqrt {{p_l}} \left| {{\alpha _l}} \right|}\\
 {\rm{s.t.}}&\ \ \sum\limits_{l = 1}^L {{p_l}}  \le P,\\
 &\ \ {p_l} \ge 0,\ \forall l, \nonumber
 \end{aligned}
 \end{equation}
 which is a convex problem. Denote by $\rho  \ge 0$ the dual variable associated with the first constraint of (P-MRT). The Lagrangian of (P-MRT) can be expressed as
 \begin{equation}
 {\cal L}\left( {\left\{ {{p_l}} \right\}_{l = 1}^L,\rho } \right) = \sum\limits_{l = 1}^L {\sqrt {{p_l}} \left| {{\alpha _l}} \right|}  - \rho \left( {\sum\limits_{l = 1}^L {{p_l}}  - P} \right).
 \end{equation}
 By applying the Karush-Kuhn-Tucker (KKT) conditions, it can be shown that the optimal solution to (P-MRT) is
 \begin{equation}\label{MRTOptimalPowerAllocation}
 p_l^ \star  = P\frac{{{{\left| {{\alpha _l}} \right|}^2}}}{{\sum\limits_{i = 1}^L {{{\left| {{\alpha _i}} \right|}^2}} }},\ \forall l.
 \end{equation}
 By substituting \eqref{MRTOptimalPowerAllocation} into \eqref{ReceivedSNRMRT2UpperBound}, \eqref{ReceivedSNRMRT2OptimalSNR} can be obtained. The proof of Proposition~\ref{ReceivedSNRMRT2OptimalUpperBound} is thus completed.
\end{appendices}


\bibliographystyle{IEEEtran}
\bibliography{refDDAM}
\end{document}